\documentclass[10pt,twocolumn,aps,prd,preprintnumbers,showpacs,superscriptaddress,nofootinbib,amsmath,amssymb,floats,floatfix,showkeys,notitlepage,longbibliography,]{revtex4-2}
\usepackage{orcidlink}
\usepackage{comment}
\usepackage{lipsum}
\usepackage{graphicx}
\usepackage{subfigure}
\usepackage{palatino}
\usepackage{sans}
\usepackage{hyperref}
\hypersetup{colorlinks=true,linkcolor=blue,urlcolor=blue,citecolor=blue}
\usepackage[toc,page]{appendix}
\usepackage[normalem]{ulem}
\usepackage{adjustbox}
\usepackage{latexsym}
\usepackage{amsmath}
\numberwithin{equation}{section}
\usepackage{amssymb}
\usepackage{amsfonts}
\usepackage{dcolumn}
\usepackage{bm}
\usepackage{tikz}
\usetikzlibrary{decorations.pathmorphing}
\usepackage{bigints}
\usepackage{array,tabularx,multirow,booktabs}
\usepackage[tracking=true]{microtype}
\usepackage{soul} 
\SetTracking{}{500}
\SetTracking{encoding={*}, shape=sc}{40}
\UseRawInputEncoding 
\allowdisplaybreaks
\usepackage[utf8]{inputenc}
\usepackage{xcolor} 
\usepackage{mathrsfs}

\newcommand{\nn}{\nonumber \\}

\begin{document} \sloppy

\title{Horizon brightened acceleration radiation from massive vector fields}

\author{Reggie C. Pantig \orcidlink{0000-0002-3101-8591}} 
\email{rcpantig@mapua.edu.ph}
\affiliation{Physics Department, School of Foundational Studies and Education, Map\'ua University, 658 Muralla St., Intramuros, Manila 1002  Philippines.}

\author{Ali \"Ovg\"un \orcidlink{0000-0002-9889-342X}}
\email{ali.ovgun@emu.edu.tr}
\affiliation{Physics Department, Faculty of Arts and Sciences, Eastern Mediterranean University, Famagusta, 99628 North Cyprus via Mersin 10, T\"urkiye.}

\begin{abstract}
In this paper, we develop a quantum-optical treatment of acceleration radiation for atoms freely falling into a Schwarzschild black hole when the ambient field is a massive spin-1 (Proca) field. Building on the HBAR framework of Scully and collaborators, we analyze two detector realizations: a charged-monopole current coupling and a physical electric-dipole coupling, both within a cavity that isolates a single outgoing Schwarzschild mode prepared in the Boulware state. Using a near-horizon stationary-phase analysis, we show that the thermal detailed-balance factor governing excitation versus absorption is universal and depends only on the near-horizon Rindler coordinate transformation. At the same time, the absolute spectra acquire distinctive Proca signatures: a hard mass threshold, polarization-dependent prefactors, and axial/polar graybody transmissions. Promoting single-pass probabilities to escaping rates yields a master equation whose steady state is geometric and whose entropy flux obeys an horizon brightened acceleration radiation-style area-entropy relation identical in form to the scalar case, with all vector-field specifics entering through the radiative area change. Our results provide a controlled pathway to probe longitudinal versus transverse responses, mass thresholds, and the role of polarization-resolved graybody transmission in acceleration radiation. More precisely, we derive the universal near-horizon kernel and show how the Proca transmission data enter the escaping probabilities, rates, and entropy flux; a dedicated numerical computation of the axial/polar graybody profiles is left for future work. This sets the stage for extensions to rotating backgrounds, alternative exterior states, and detector-engineering strategies.
\end{abstract}

\date{\today}
\maketitle

\section{Introduction}\label{sec1}
Black holes provide a unique meeting point between gravitation, quantum field theory, and statistical physics. Seminal work showed that quantum fields in curved spacetime endow black holes with a thermal character, leading to particle emission and an associated thermodynamics governed by horizon geometry \cite{Bekenstein:1973ur, Hawking:1974rv, Hawking:1975vcx}. The modern picture originates with Hawking's prediction of black‐hole radiation and the area-entropy paradigm that followed, which together frame today's discussions of information, equilibration, and backreaction in strong gravity.

Beyond spontaneous emission at future null infinity, accelerated motion in vacuum can itself stimulate radiation. In flat spacetime this appears as the Unruh effect, for which accelerated detectors register a thermal spectrum set by their proper acceleration; comprehensive accounts have clarified both foundational aspects and operational detector models \cite{Unruh:1976db}. In black‐hole spacetimes, the same near‐horizon mapping that underlies Hawking's result organizes the response of localized probes and the structure of physically relevant vacuum states \cite{Boulware:1974dm}

A particularly incisive formulation of this physics, which is directly relevant for quantum‐optical probes, is the horizon brightened acceleration radiation (HBAR) program introduced by Scully and collaborators \cite{Scully:2017utk}. In that setting, atoms prepared in their ground state fall through a cavity toward a Schwarzschild horizon while the exterior field is in the Boulware vacuum; the atoms emit acceleration radiation via a counterrotating channel, and a master‐equation treatment yields an entropy flux that closely parallels black‐hole thermodynamic relations. Subsequent work developed this quantum‐optics viewpoint and emphasized its complementarity to conventional Hawking derivations \cite{Azizi:2021qcu,Azizi:2021yto}. Since then, several studies have emerged to use accelerated radiation to probe certain geometric signatures from different black hole models \cite{Svidzinsky:2018jkp,Ben-Benjamin:2019opz,Azizi:2020gff,Azizi:2021qcu,Azizi:2021yto,Camblong:2020pme,Sen:2022cdx,Sen:2022tru,Das:2023rwg,Das:2025rzz,Jana:2024fhx,Jana:2025hfl,Bukhari:2022wyx,Bukhari:2023yuy,Ovgun:2025isv,Pantig:2025okn,Pantig:2025igg}, and even on time-dependent black hole metrics \cite{Dalui:2020qpt,Kaplanek:2021sbo,Shallue:2025zto,Clarke:2024lwi,Carullo:2025oms}.

Motivated by both conceptual and phenomenological considerations, we revisit this scenario for a massive spin-1 field governed by the Proca equations. Massive vector fields introduce qualitatively new ingredients: a longitudinal polarization with distinct near-horizon couplings; a hard spectral threshold set by the rest mass that reshapes low-frequency response; and polarization-dependent propagation through curvature-induced barriers that control the greybody transmission to infinity. These features are well studied in black hole perturbation theory, where the Proca field separates into axial and polar sectors with one and two effective channels, respectively, and displays characteristic quasinormal and quasibound spectra \cite{Konoplya:2005hr,Rosa:2011my,Fernandes:2021qvr}. They also connect to broader interest in hidden photons and dark sectors, for which black hole environments provide sensitive laboratories.

From the propagation side, the Schwarzschild problem for Proca fields has been analyzed in detail, including decoupled master equations, axial/polar structure, and numerical spectra; recent work has begun to systematize greybody factors for massive vectors, underscoring the importance of transmission coefficients in any observational or analogue setup \cite{Rosa:2011my}. {\color{black}These developments make the Proca generalization of the HBAR framework both timely and natural. The physical interest is not merely that one obtains a spin-1 version of the scalar calculation, but that massive vectors introduce qualitatively new observables: a longitudinal polarization absent in scalar and Maxwell cases, a hard spectral threshold at \(\nu=m_V\), and polarization-resolved propagation through axial/polar Schwarzschild barriers. Within a cavity-selected HBAR protocol, these features imply a distinctive \textit{gap-plus-tail} spectrum, a channel-dependent turn-on near threshold, and detector-dependent sensitivity to longitudinal versus transverse response. These are semirealistic signatures in the sense relevant to detector theory and analogue implementations: they specify which observable features would distinguish a massive vector field from scalar or massless spin-1 radiation in any mode-resolved measurement. More broadly, because light vectors and hidden photons are actively studied in black hole environments, the Proca case provides a useful analytic template for separating universal near-horizon thermality from field-specific spin-1 spectral structure.} More broadly, interest in light vectors around black holes, especially in the context of ultralight hidden photons, has emphasized superradiant dynamics and observational signatures, underscoring the phenomenological value of polarization- and mass-threshold effects in spin-1 spectra \cite{Brito:2015oca,Baryakhtar:2017ngi}. Finally, the role of cavities and localized detectors in curved backgrounds has been analyzed in detector-theory language, offering complementary perspectives on how boundary conditions and motion shape response, which are concepts directly mirrored by the cavity protocol adopted here \cite{Ahmadzadegan:2013iua}.

{\color{black}We adopt Scully \textit{et al.}'s \cite{Scully:2017utk} kinematic and operational scaffold (radial free fall into a Schwarzschild black hole, Boulware exterior state, and a mode-selecting cavity) to isolate an outgoing channel and evaluate the detector response. This choice should be understood operationally rather than as a claim about the globally physical late-time state of an evaporating black hole. In the present HBAR protocol, the role of the Boulware vacuum is to provide a zero-occupation reference state for the cavity-filtered outgoing mode, so that the atom's excitation isolates the acceleration-radiation channel without contamination from a preexisting thermal bath. Our analysis keeps the detector modeling explicit: we treat (i) a charged-monopole current coupling and (ii) a physical electric-dipole coupling to the local field strength, thereby bracketing the range of realistic atom-field interactions in this geometry. {\color{black}Throughout, we emphasize what is universal (the near-horizon phase structure and the resulting detailed balance) versus what is model-dependent (polarization projectors, mass thresholds, and greybody transmission). In the present work, the greybody sector is incorporated at the level of general Schwarzschild scattering data: we identify the transmission coefficients that enter the cavity-selected HBAR rates, but we do not attempt a separate numerical computation of the axial/polar Proca greybody profiles. Thus, our main result is a factorized analytic framework in which the local near-horizon kernel is derived explicitly, while the global propagation sector enters through transmission inputs.}

We stress, however, that the Boulware state is an idealized baseline for the cavity-selected problem. For an evaporating black hole, the more physical exterior state is the Unruh vacuum, whereas the Hartle-Hawking state describes thermal equilibrium. In those alternatives, the same near-horizon worldline kernel derived below remains the local building block, but the selected mode is no longer initially empty, so the effective emission/absorption rates acquire additional state-dependent Bose factors. The present paper therefore isolates the vacuum-seeded HBAR channel; a systematic incorporation of Unruh/Hartle-Hawking occupation effects is left for future work.}

We organize the paper as follows: Sec. \ref{sec2} fixes the physical setting and kinematics of radial infall, the exterior vacuum, and frequency labels. Sec. \ref{sec3} formulates the Proca dynamics on Schwarzschild, including separation into axial/polar sectors and mode normalization. Sec. \ref{sec4} specifies the two detector models and derives the corresponding worldline amplitudes. Sec. \ref{sec5} evaluates the excitation probability using a near-horizon stationary-phase analysis, isolating the universal thermal kernel and identifying the Proca-specific prefactors without invoking explicit formulas here. Sec. \ref{sec6} incorporates the cavity protocol and greybody transmission to promote probabilities to emission/absorption rates and to a master equation for the mode occupations. Sec. \ref{sec7} develops the entropy-flux analysis (HBAR) for massive spin-1 quanta and relates it to the black hole area change. Sec. \ref{sec8} assembles consistency checks and theoretical subtleties, including control of the large-gap expansion, the role of the Proca constraint in place of gauge freedom, and the connection to Unruh/DeWitt detector logic. We close in Sec. \ref{sec9} with a brief summary and outlook.

\section{Preliminaries and Conventions} \label{sec2}
We consider two-level atoms falling radially from rest at infinity into a nonrotating Schwarzschild black hole of mass \(M\). The exterior quantum field is prepared in the Boulware vacuum. A mode-selecting cavity isolates ounterpropagating outgoing modes so that the atom-field interaction reduces to a single relevant frequency label \(\nu\) for distant observers. Kinematics is encoded by the Schwarzschild line element and the tortoise coordinate \(r_*\), and the atomic worldline is specified by \(t(\tau)\) and \(r(\tau)\) for proper time \(\tau\). The near-horizon redshift makes the WKB phase of outgoing modes depend on the combination \(t-r_*\), which will be the driver of the thermal detailed-balance factor later on.

\subsection{Geometry, trajectory, and coordinates} \label{ssec2.1}
We work outside the Schwarzschild radius \(r_g\equiv 2GM/c^2\) with line element \cite{Schwarzschild:1916uq,Chandrasekhar:1985kt}
\begin{equation}
ds^2=f(r)\,c^2dt^2-\frac{dr^2}{f(r)}-r^2(d\theta^2+\sin^2\theta\,d\phi^2), \label{2.1}
\end{equation}
with $f(r)=1-r_g/r$. For wave propagation, it is convenient to introduce the Regge-Wheeler tortoise coordinate \cite{Regge:1957td,Fiziev:2007es}
\begin{equation}
r_*(r)=r+r_g\ln\!\left(\frac{r}{r_g}-1\right), \label{2.2}
\end{equation}
so that outgoing high-frequency modes behave locally as \(e^{-i\nu(t-r_*/c)}\) near the horizon (the precise field content will be specified later). In a small neighborhood of the horizon, one may use the Rindler approximation obtained by setting \(r=r_g+z^2/(4r_g)\) with \(0<z\ll r_g\), which yields \cite{Birrell:1982ix}
\begin{equation}
ds^2\simeq \frac{z^2}{4r_g^2}\,c^2dt^2-dz^2-r_g^2 d\Omega_2^2, \label{2.3}
\end{equation}
making explicit the uniform proper acceleration of static worldlines at fixed \(z\). 

Following Ref. \cite{Scully:2017utk}, we adopt the dimensionless variables
\begin{align}
r&\rightarrow r_g\,r,\nn t&\rightarrow (r_g/c)\,t,\nn \omega&\rightarrow (c/r_g)\,\omega,\nn \nu&\rightarrow (c/r_g)\,\nu, \label{2.4}
\end{align}
and consider a test atom released from rest at infinity to fall radially. Henceforth, all coordinates and frequencies are treated as dimensionless quantities scaled by $r_g$ and $c$, unless otherwise noted. To continue, the radial geodesic first integrals then read
\begin{equation}
\;\frac{dr}{d\tau}=-\frac{1}{\sqrt{r}},\qquad \frac{dt}{d\tau}=\frac{r}{r-1}\;, \label{2.5}
\end{equation}
where \(t\) is the Schwarzschild time and \(\tau\) the atom's proper time [both dimensionless in the convention of \eqref{2.4}]. Integrating \eqref{2.5}, we obtain the parametric trajectory
\begin{equation}
\tau(r)=-\frac{2}{3}\,r^{3/2}+{\rm const}, \label{2.6}
\end{equation}
\begin{equation}
t(r)=-\frac{2}{3}\,r^{3/2}-2\sqrt{r}-\ln\!\left(\frac{\sqrt{r}-1}{\sqrt{r}+1}\right)+{\rm const}, \label{2.7}
\end{equation}
and the corresponding tortoise coordinate \cite{Regge:1957td,Fiziev:2007es}
\begin{equation}
r_*(r)=r+\ln(r-1). \label{2.8}
\end{equation}
Equations \eqref{2.5}--\eqref{2.8} completely specify \(t(\tau)\), \(r(\tau)\), and \(r_*(\tau)\) for the ensuing stationary-phase analysis of the atom-field matrix elements. These expressions coincide with the kinematic relations used in Appendixes A and B of Scully \textit{et al.} \cite{Scully:2017utk}, to which we will refer for consistency of conventions. 

We remark that the near-horizon Rindler form \eqref{2.3} rationalizes the universality of the phase \(t-r_*\) and will underpin the Planckian factor arising later from the \(\tau\)-integral along the worldline. 2) The choice of additive constants in \eqref{2.6}--\eqref{2.7} fixes the zero of \(t\) and \(\tau\) and is immaterial for transition probabilities, which depend only on phase differences. 

\subsection{Field state and cavity} \label{ssec2.2}
We take the exterior quantum field to be in the Boulware vacuum, i.e. the vacuum defined by positive Killing frequency with respect to \(\partial_t\) in the static exterior. Operationally, we place a one-dimensional, axially symmetric cavity along the radial direction to isolate a single, outgoing mode counterpropagating with the infalling atom cloud (the same role played by the cavity/mode selector in Scully et al \cite{Scully:2017utk}). The cavity suppresses contamination by incoming flux and renders the coarse-grained master-equation description well-posed; leakage can be neglected provided the photon loss rate is small compared to the atom-induced absorption rate.  

We restrict modes to a finite interval of the tortoise coordinate,
\begin{equation}
r_*^{(b)}< r_* < r_*^{(t)}, \label{2.9}
\end{equation}
with the bottom at \(r_b>r_g\) and the top at \(r_t<\infty\). Reflecting boundary conditions at \(r_b\,r_t\) (perfect-mirror idealization) select a discrete set of standing waves. The physically relevant result is obtained by the standard limiting procedure
\begin{equation}
r_b \to r_g^{+},\qquad r_t \to \infty, \label{2.10}
\end{equation}
which reproduces the continuum of exterior modes while preserving the single-mode filtering implemented inside the cavity. This is precisely the setup advocated to reach the thermal steady state of the field mode and to connect with the density-matrix analysis. 

In the Boulware state, all annihilation operators \(a_{\nu,\lambda}\) associated with outgoing positive-\(t\)-frequency modes annihilate the vacuum,
\begin{equation}
a_{\nu,\lambda}\,|0_B\rangle=0,\qquad \nu>0, \label{2.11}
\end{equation}
where \(\lambda\) labels polarization (for the Proca field) or the single scalar polarization \cite{Boulware:1974dm,Candelas:1980zt}.  {\color{black}This state choice is best viewed as an operational preparation of the cavity-filtered outgoing mode, not as a claim that the global exterior state of an evaporating black hole is Boulware. The purpose of the cavity is precisely to isolate a single outgoing channel with vanishing initial occupation, so that the atom's counterrotating excitation probes the acceleration-radiation mechanism in its cleanest form. In this sense, the present calculation generalizes the original HBAR protocol to the Proca case.

By contrast, in the Unruh vacuum the outgoing sector carries a thermal occupation while the ingoing sector remains in vacuum; in the Hartle-Hawking vacuum both sectors are thermally populated. For those states, the near-horizon phase structure and the associated kernel derived in Secs. \ref{sec5}--\ref{sec8} remain unchanged, but the cavity-mode master equation would acquire additional bath-dependent stimulated-emission and induced-absorption terms proportional to the corresponding occupation numbers. Thus, the Boulware calculation should be read as isolating the vacuum contribution to the selected outgoing channel, while alternative exterior states add state-dependent occupation effects on top of the same local kinematic kernel.}

The near-horizon WKB form of each outgoing mode depends on the universal phase \(t-r_*\); hence, when the atom falls through the cavity, the dominant contribution to the transition amplitude is governed by this phase, while the cavity merely enforces mode selection and provides the coarse graining needed for a laser-style master equation and the ensuing HBAR entropy analysis.

\subsection{Frequency labels and asymptotic kinematics} \label{ssec2.3}

We label modes by the Killing frequency \(\nu>0\) measured by distant inertial observers. For the Proca field with rest mass \(m_V\), the asymptotic (flat-space) dispersion relation fixes the radial wave number and group velocity,
\begin{equation}
k_\infty(\nu)=\sqrt{\nu^2-m_V^2},\qquad v_g(\nu)=\frac{k_\infty(\nu)}{\nu}, \label{2.12}
\end{equation}
so only \(\nu\ge m_V\) contribute to propagating flux at \(\mathscr{I}^+\). Local measurements differ because of gravitational redshift. A static observer at radius \(r\) registers the proper frequency
\begin{equation}
\nu_{\text{stat}}(r)=\frac{\nu}{\sqrt{f(r)}}, \qquad f(r)=1-\frac{1}{r}, \label{2.13}
\end{equation}
and the locally measured 3-momentum satisfies the Proca dispersion \(\nu_{\text{stat}}^2=k_{\text{stat}}^2+m_V^2\).

By contrast, the infalling atom samples the comoving frequency given by the Doppler-gravitational invariant \cite{Carroll_2019}
\begin{equation}
\Omega(\tau,\nu)= -\,k_\mu u^\mu = \nu\!\left(\frac{dt}{d\tau}-\frac{dr_*}{d\tau}\right), \label{2.14}
\end{equation}
where \(u^\mu\) is the atom's four-velocity and we have used the outgoing WKB phase \(\sim e^{-i\nu(t-r_*)}\) near the horizon. We choose the associated covector \(k_\mu = -\partial_\mu (t - r_*) = (-1,\,+\partial_r r_*,\,0,\,0)
      = \left(-1,\,(1-1/r)^{-1},\,0,\,0\right)\),
so that \(k_t = -\nu\), \(k_r = \nu \, dr_*/dr\). Using the geodesic components $u^t = r/(r-1)$ and $u^r = -1/\sqrt{r}$ from \eqref{2.5}, the invariant contraction \eqref{2.14} yields

\begin{align}
\Omega(\tau,\nu)
&\equiv -\,k_\mu u^\mu
= \nu\!\left(\frac{dt}{d\tau} - \frac{dr_*}{d\tau}\right) \nn
&= \nu\!\left(\frac{r}{r-1} + \frac{\sqrt{r}}{r-1}\right)
= \nu\,\frac{r+\sqrt{r}}{r-1}\,,
\label{2.15}
\end{align}
which diverges as $r\to1^+$ and rationalizes the dominance of the near-horizon segment in the asymptotic evaluation of transition amplitudes.

\begin{figure}
    \centering
    \includegraphics[width=\columnwidth]{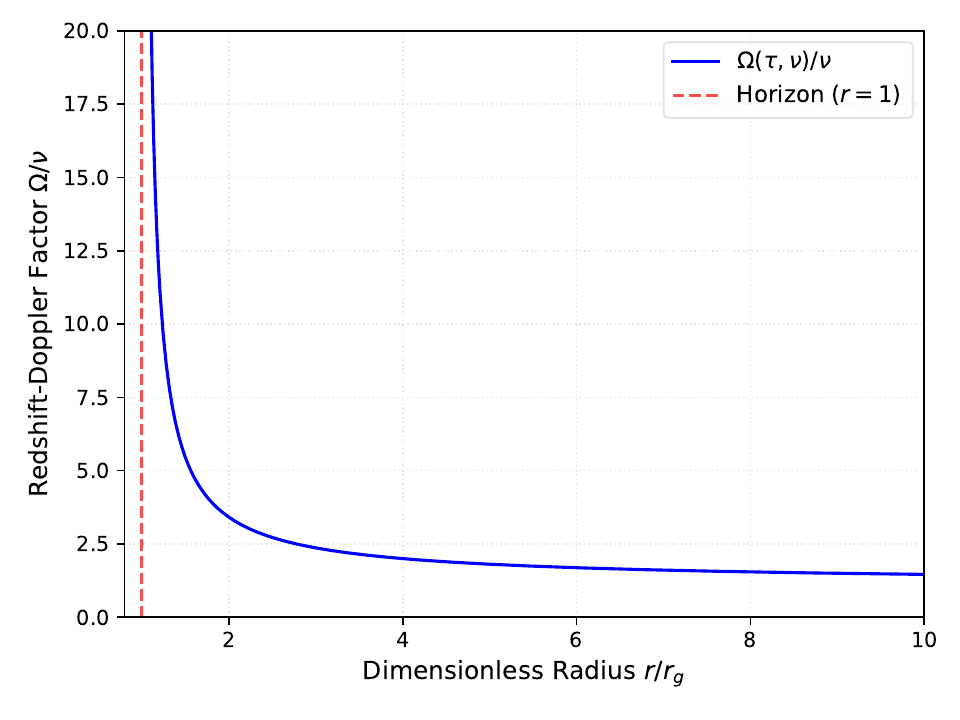}
    \caption{The comoving redshift-Doppler factor $\Omega/\nu$ plotted against the dimensionless radial coordinate $r$. The vertical dashed line represents the event horizon.}
    \label{fig1}
\end{figure}

The kinematic origin of the excitation is visualized in Fig. \ref{fig1}, which plots the comoving frequency ratio $\Omega(\tau, \nu)/\nu$ as a function of the radial distance. As the atom approaches the Schwarzschild radius ($r \to 1^+$), the redshift-Doppler factor grows without bound. This divergence or horizon brightening implies that the infalling atom perceives the finite-frequency vacuum modes of the laboratory frame as having arbitrarily high energies. Consequently, the atom can undergo transitions mediated by the counterrotating terms in the interaction Hamiltonian, which are energetically forbidden in inertial frames but become dominant near the horizon.

Polarization labels will be denoted by \(\lambda=1,2,3\). At infinity, these correspond to three physical Proca polarizations; near the black hole, they organize into axial/polar sectors of the separated field equations. For later use, we introduce the standard polarization projector
\begin{equation}
\Pi_{\mu\nu}(\nu,m_V)\equiv \sum_{\lambda}\varepsilon^{(\lambda)}_\mu(\nu)\,\varepsilon^{(\lambda)}_\nu(\nu), \label{2.16}
\end{equation}
so that matrix elements for the charged-monopole and dipole detector models reduce to scalar contractions with \(u^\mu\) or the atom's instantaneous spatial direction, respectively. The kinematic content of this subsection is thus: (i) a hard threshold at \(\nu=m_V\), (ii) an asymptotic phase space governed by \(k_\infty(\nu)\), and (iii) a large, universal blueshift of the comoving frequency \(\Omega(\tau,\nu)\) dictated by \eqref{2.15}. These ingredients will control the prefactors of the excitation probabilities, while the Planckian detailed-balance factor will originate from the phase structure implicit in \eqref{2.14}--\eqref{2.15}.

\section{Proca field on Schwarzschild} \label{sec3}
We formulate a massive spin-1 field on the Schwarzschild exterior as a Proca vector \(A_\mu\) with field strength \(F_{\mu\nu}\). The key structural features are: (i) a mass term that removes gauge redundancy and enforces the covariant Lorenz constraint \(\nabla_\mu A^\mu=0\); (ii) equations of motion that reduce, in Ricci-flat backgrounds, to a vector wave equation with a mass term; and (iii) three physical polarizations. These ingredients will permit a clean separation into axial/polar sectors and a near-horizon WKB analysis in later subSecs., where the outgoing mode phase aligns with the convention fixed in Sec. \ref{sec2}, \(e^{-i\nu(t-r_*)}\).

\subsection{Equations and constraints} \label{ssec3.1}
We work on the exterior Schwarzschild geometry already defined in Sec. \ref{sec2}. The Proca action is \cite{Rosa:2011my}
\begin{align}
S[A]&=\int\! d^4x\,\sqrt{-g}\left(
-\frac{1}{4}\,F_{\mu\nu}F^{\mu\nu}
+\frac{m_V^{2}}{2}\,A_\mu A^\mu\right), \nn
F_{\mu\nu}&\equiv \nabla_\mu A_\nu-\nabla_\nu A_\mu . \label{3.1}
\end{align}
Variation with respect to \(A_\mu\) yields the Proca equations of motion \cite{Rosa:2011my}
\begin{equation}
\nabla_\nu F^{\mu\nu}+m_V^2 A^\mu=0. \label{3.2}
\end{equation}
Taking the covariant divergence of \eqref{3.2} gives the Proca constraint
\begin{equation}
m_V^2\,\nabla_\mu A^\mu=0\quad\Longrightarrow\quad \;\nabla_\mu A^\mu=0\;, \label{3.3}
\end{equation}
which replaces gauge freedom: there is no gauge symmetry for \(m_V\neq 0\), and \(\nabla\!\cdot\!a=0\) follows dynamically.

It is often convenient to eliminate \(F_{\mu\nu}\) using \eqref{3.3}. Acting with \(\nabla_\nu\) on the definition of \(F_{\mu\nu}\) and using the commutator of covariant derivatives, one obtains the equivalent second-order vector equation \cite{Birrell:1982ix,Carroll_2019}
\begin{equation}
\left(\Box\,\delta^\mu_{\ \nu}-R^\mu_{\ \nu}+m_V^2\,\delta^\mu_{\ \nu}\right)A^\nu=0,
\qquad \Box\equiv \nabla_\alpha\nabla^\alpha. \label{3.4}
\end{equation}
In the Schwarzschild exterior \(R_{\mu\nu}=0\), so \eqref{3.4} simplifies to
\begin{equation}
\left(\Box+m_V^2\right)A^\mu=0\quad\text{with}\quad \nabla_\mu A^\mu=0. \label{3.5}
\end{equation}
Equations \eqref{3.2}--\eqref{3.5} show that there are three propagating degrees of freedom, corresponding to the massive vector's physical polarizations. For later use, we also record the symmetric stress-energy tensor obtained from \eqref{3.1},
\begin{align}
T_{\mu\nu}&= F_{\mu\alpha}F_{\nu}^{\ \alpha}
-\frac{1}{4}\,g_{\mu\nu}F_{\alpha\beta}F^{\alpha\beta}
+m_V^2\nn
&\times \!\left(A_\mu A_\nu-\frac{1}{2}g_{\mu\nu}A_\alpha A^\alpha\right), \label{3.6}
\end{align}
which is conserved, \(\nabla^\mu T_{\mu\nu}=0\), when \eqref{3.2} holds \cite{Birrell:1982ix}.

The dynamical pair \((A_\mu\,\nabla\!\cdot\!a=0)\) will be our starting point for separation of variables and for defining polarization sums and mode normalizations in the Schwarzschild exterior.

\subsection{Master radial equations and effective potentials} \label{ssec3.2}
We separate the Proca field \(A_\mu\) on the Schwarzschild exterior [with \(f(r)=1-1/r\) and tortoise \(dr_*/dr=f^{-1}\)] into vector spherical harmonics. The decomposition splits into an axial (odd-parity) sector with one master variable and a polar (even-parity) sector carrying two coupled master variables. Throughout we use the positive-frequency convention for outgoing modes fixed in Sec. \ref{sec2}, namely phases \(e^{-i\nu(t-r_*)}\).

For each \(\ell\ge1\) and \(m\), the axial sector is governed by a single master function \(\Psi_{\ell}^{\text{(ax)}}(r)\) satisfying a Schr\"odinger-type radial equation
\begin{equation}
\frac{d^2 \Psi_{\ell}^{\text{(ax)}}}{dr_*^{2}}
+\left[\nu^{2}-V_{\ell}^{\text{(ax)}}(r)\right]\Psi_{\ell}^{\text{(ax)}}=0, \label{3.7}
\end{equation}
with the effective potential
\begin{equation}
\;V_{\ell}^{\text{(ax)}}(r)=f(r)\!\left[\frac{\ell(\ell+1)}{r^{2}}+m_V^{2}\right].\; \label{3.8}
\end{equation}
Two key limits are immediate: near the horizon \(r\to1^{+}\), \(f\to0\) so \(V_{\ell}^{\text{(ax)}}\to0\); at infinity \(V_{\ell}^{\text{(ax)}}\to m_V^{2}+\mathcal{O}(r^{-2})\). Hence, asymptotic waves have wave number \(k_\infty=\sqrt{\nu^{2}-m_V^{2}}\) and the threshold \(\nu\ge m_V\).

The polar (even) sector contains the remaining two physical polarizations of the massive field. After eliminating the algebraic constraint \(\nabla_\mu A^\mu=0\) and metric connections, one arrives at a coupled system for two master functions \(\Phi_{\ell}^{(1)}(r)\) and \(\Phi_{\ell}^{(2)}(r)\):
\begin{equation}
\frac{d^{2}}{dr_*^{2}}
\begin{pmatrix}
\Phi_{\ell}^{(1)} \\ \Phi_{\ell}^{(2)}
\end{pmatrix}
+\left[\nu^{2}\mathbf{I}-\mathbf{U}_{\ell}(r)\right]
\begin{pmatrix}
\Phi_{\ell}^{(1)} \\ \Phi_{\ell}^{(2)}
\end{pmatrix}=0, \label{3.9}
\end{equation}
where the \(2\times2\) potential matrix has the general structure
\begin{equation}
\mathbf{U}_{\ell}(r)=
f(r)\!
\begin{pmatrix}
\displaystyle m_V^{2}+\frac{\ell(\ell+1)}{r^{2}}+ \frac{\alpha_{\ell}(r)}{r^{3}}
&
\displaystyle \frac{\beta_{\ell}(r)}{r^{2}} \\
\displaystyle \frac{\gamma_{\ell}(r)}{r^{2}}
&
\displaystyle m_V^{2}+\frac{\ell(\ell+1)}{r^{2}}+\frac{\delta_{\ell}(r)}{r^{3}}
\end{pmatrix}. \label{3.10}
\end{equation}
The functions \(\alpha_\ell,\beta_\ell,\gamma_\ell,\delta_\ell\) are rational in \(r\) (dimensionless in our units) and arise from curvature couplings and the Proca constraint; they fall at least as \(r^{-1}\). Two facts suffice for our purposes: (1) Since \(\mathbf{U}_\ell\propto f(r)\), each entry vanishes linearly as \(r\to1^{+}\). Thus, both polar master fields share the same plane wave behavior as the axial field:
\begin{equation}
   \Phi_{\ell}^{(i)}\sim e^{-i\nu(t-r_*)},\qquad r\to1^{+}, \ i=1,2. \label{3.11}
\end{equation}
This is the origin of the spin-independence of the near-horizon thermal factor used later. (2) As \(r\to\infty\), \(\mathbf{U}_\ell(r)\to \left(m_V^{2}+\ell(\ell+1)/r^{2}\right)\mathbf{I}+\mathcal{O}(r^{-3})\). Hence both polar modes propagate with the same asymptotic wave number \(k_\infty=\sqrt{\nu^{2}-m_V^{2}}\) and the same mass threshold \(\nu\ge m_V\).

It is convenient (though not necessary for the leading WKB analysis) to diagonalize \(\mathbf{U}_\ell\) by a local \(r\)-dependent rotation\, yielding two decoupled master variables \(\Psi_{\ell}^{\text{(pol)}\pm}\) obeying
\begin{equation}
\frac{d^2 \Psi_{\ell}^{\text{(pol)}\pm}}{dr_*^{2}}
+\left[\nu^{2}-V_{\ell}^{\text{(pol)}\pm}(r)\right]\Psi_{\ell}^{\text{(pol)}\pm}=0, \label{3.12}
\end{equation}
with effective potentials \(V_{\ell}^{\text{(pol)}\pm}(r)\) that share the same limiting behaviors as in \eqref{3.8}, viz.
\begin{equation}
V_{\ell}^{\text{(pol)}\pm}(r)= f(r)\!\left[m_V^{2}+\frac{\ell(\ell+1)}{r^{2}}\right] + \mathcal{O}\!\left(\frac{f(r)}{r^{3}}\right). \label{3.13}
\end{equation}
The \(\mathcal{O}(f/r^{3})\) terms encode modest polarization-dependent barrier differences (greybody effects) that will only modify prefactors of excitation probabilities.

For any master equation of the form \(d^{2}\Psi/dr_*^{2}+[\nu^{2}-V(r)]\Psi=0\), the Wronskian
\begin{equation}
\mathcal{W}=\Psi^\ast \frac{d\Psi}{dr_*}-\frac{d\Psi^\ast}{dr_*}\Psi \label{3.14}
\end{equation}
is constant in \(r_*\). We normalize solutions so that a unit outgoing flux at infinity corresponds to \(\Psi\sim e^{-ik_\infty r_*}\) with \(k_\infty=\sqrt{\nu^{2}-m_V^{2}}\). Near the horizon the same mode behaves as \(e^{-i\nu(t-r_*)}\), consistent with Sec. \ref{sec2}. These conventions set the stage for the stationary-phase worldline integrals and for introducing polarization sums and greybody factors in later Secs..

\subsection{Mode normalization and density of states} \label{ssec3.3}
We expand the Proca field in positive-frequency, outgoing Schwarzschild modes labeled by \((\nu,\ell,m,\lambda)\),
\begin{align}
A_\mu(x)&=\sum_{\ell m}\sum_{\lambda=1}^{3}\int_{m_V}^{\infty}\! d\nu\;
\Big[a_{\nu\ell m\lambda}\,\mathcal{A}^{(\lambda)}_{\mu;\nu\ell m}(x) \nn
&+a_{\nu\ell m\lambda}^\dagger\,\mathcal{A}^{(\lambda)\,*}_{\mu;\nu\ell m}(x)\Big], \label{3.15}
\end{align}
where the mode functions are normalized by unit outgoing flux at infinity. As \(r\to\infty\), each mode takes the form
\begin{align}
    \mathcal{A}^{(\lambda)}_{\mu;\nu\ell m}(x)&\sim\frac{1}{r}\frac{\varepsilon^{(\lambda)}_{\mu}(\nu,\hat{\mathbf{k}})}{\sqrt{k_\infty(\nu)}}Y_{\ell m}(\theta,\phi) \, e^{-i\nu t}e^{-ik_\infty(\nu) r_*},\nn k_\infty(\nu)&=\sqrt{\nu^2-m_V^2}. \label{3.16}
\end{align}
so that the associated radial Wronskian, with the \(1/\sqrt{k_\infty}\bullet\) normalization, is \(\mathcal W=-2i\) and the outgoing flux at infinity is unity. The three polarization vectors satisfy the standard massive-vector orthogonality; asymptotically (in the locally flat frame) the polarization sum is
\begin{equation}
\sum_{\lambda=1}^{3}\varepsilon^{(\lambda)}_{\mu}\,\varepsilon^{(\lambda)}_{\nu}
=-\eta_{\mu\nu}+\frac{k_\mu k_\nu}{m_V^{2}}, \label{3.17}
\end{equation}
with \(k^\mu=(\nu,\mathbf{k})\) and \(\eta_{\mu\nu}\) the Minkowski metric with signature \( (+,-\,-\,-) \).

To connect with the cavity description in Sec. \ref{sec2}, we place reflecting walls at tortoise locations \(r_*^{(b)}<r_*<r_*^{(t)}\) of length \(L_*\equiv r_*^{(t)}-r_*^{(b)}\). Standing-wave conditions quantize the asymptotic wave number as \(k_\infty L_*=\pi n\) (integer \(n\)). In the continuum limit \(L_*\to\infty\) with a narrow frequency window, the number of modes in \([\nu,\nu+d\nu]\) for a fixed \((\ell,m,\lambda)\) is
\begin{equation}
\;\rho(\nu)\,d\nu=\frac{L_*}{\pi}\,\frac{dk_\infty}{d\nu}\,d\nu
=\frac{L_*}{\pi}\,\frac{\nu}{\sqrt{\nu^{2}-m_V^{2}}}\,\Theta(\nu-m_V)\,d\nu\;, \label{3.18}
\end{equation}
which we will use as the replacement \(\sum_n\to\int d\nu\,\rho(\nu)\) when converting discrete sums to integrals. In practice, our single-mode selection absorbs \(L_*\) into the effective coupling and loss parameters; what remains dynamically relevant is that (i) the flux normalization fixes the asymptotic amplitude via the Wronskian, and (ii) the asymptotic phase space is governed by \(k_\infty(\nu)\) and the sharp threshold \(\nu\ge m_V\).

Near the horizon, all master variables in Sec. \ref{ssec3.2} reduce to plane waves with phase \(e^{-i\nu(t-r_*)}\); the normalization at infinity then propagates inward by constancy of the Wronskian. These conventions ensure that transition amplitudes derived along the infalling worldline carry the correct polarization sum \eqref{3.17} and the correct kinematic weight through \(k_\infty(\nu)\) and \(\rho(\nu)\) in \eqref{3.18}.

\section{Detector models and atom-field interaction} \label{sec4}
We model the atom as a two-level system with gap \(\omega>0\) moving on the infalling geodesic specified in Sec. \ref{sec2}. The Proca field is expanded in properly normalized outgoing Schwarzschild modes as in Sec. \ref{sec3}. Interaction is treated perturbatively in the coupling constants. We consider two couplings: a charged-monopole (current) coupling, where the atomic monopole couples to \(u^\mu A_\mu\) along the trajectory, and a physical dipole coupling to the local electric field. In each case, excitation of a ground-state atom in the Boulware vacuum is mediated by the counterrotating channel (atom excitation + field emission). The resulting amplitudes reduce to a single worldline integral with the near-horizon phase \(\nu[t(\tau)-r_*(\tau)]\).

\subsection{Charged-monopole (current) coupling} \label{ssec4.1}
We take the interaction Lagrangian density along the worldline to be
\begin{equation}
L_{\text{int}}(\tau)= g_V\,m(\tau)\,u^\mu(\tau)A_\mu\!\left(x(\tau)\right), \label{4.1}
\end{equation}
where \(g_V\) is a small real coupling, \(u^\mu\) the atom's four-velocity, and \(m(\tau)\) the atomic monopole operator (the reader is cautioned to distinguish this from the vector mass $m_V$, and the magnetic quantum number $m$). In the atom's interaction picture,
\begin{equation}
m(\tau)= m_{eg}\,e^{+i\omega \tau}\,\sigma_{eg}+ m_{ge}\,e^{-i\omega \tau}\,\sigma_{ge}, \quad m_{ge}=m_{eg}^\ast, \label{4.2}
\end{equation}
with \(\sigma_{eg}=|e\rangle\langle g|\). The Proca field expansion in outgoing, positive-frequency Schwarzschild modes with unit outgoing flux at infinity is
\begin{align}
A_\mu(x)&=\sum_{\ell m}\sum_{\lambda=1}^{3}\int_{m_V}^{\infty}\! d\nu\;
\Big[a_{\nu\ell m\lambda}\,\mathcal{A}^{(\lambda)}_{\mu;\nu\ell m}(x) \nn
&+a_{\nu\ell m\lambda}^\dagger\,\mathcal{A}^{(\lambda)\,*}_{\mu;\nu\ell m}(x)\Big], \label{4.3}
\end{align}
with asymptotics
\begin{align}
\mathcal{A}^{(\lambda)}_{\mu;\nu\ell m}(x)&\sim\frac{1}{r}\,
\frac{\varepsilon^{(\lambda)}_{\mu}(\nu\,\hat{\mathbf{k}})}{\sqrt{k_\infty(\nu)}}\,Y_{\ell m}(\theta,\phi)\,
e^{-i\nu t}\,e^{-ik_\infty(\nu) r_*}, \nn k_\infty(\nu)&=\sqrt{\nu^2-m_V^2}. \label{4.4}
\end{align}
Near the horizon each outgoing mode reduces to the universal WKB form \(e^{-i\nu[t-r_*]}\), consistent with \eqref{2.15} and the sign convention fixed earlier.

Starting from the ground state \(|g;0_B\rangle\), the only first-order process that excites the atom in the Boulware vacuum is the counterrotating one, \(|g;0_B\rangle\to |e;1_{\nu\ell m\lambda}\rangle\). The corresponding amplitude for a selected cavity mode \((\nu,\ell,m,\lambda)\) is
\begin{equation}
\mathcal{A}^{\text{(mono)}}_{\nu\ell m\lambda}
= -\,i g_V m_{eg}\!\int_{-\infty}^{+\infty}\! d\tau\;
e^{+i\omega \tau}\,u^\mu(\tau)\,
\mathcal{A}^{(\lambda)\,*}_{\mu;\nu\ell m}\!\left(x(\tau)\right). \label{4.5}
\end{equation}
Inserting \eqref{4.4} and restricting to the outgoing branch selected by the cavity, we obtain a one-dimensional integral of the form
\begin{align}
\mathcal{A}^{\text{(mono)}}_{\nu\ell m\lambda}
&= -\,i g_V m_{eg}\!\int d\tau\;
\frac{u^\mu(\tau)\,\varepsilon^{(\lambda)}_{\mu}(\nu\,\hat{\mathbf{k}})}{\sqrt{k_\infty(\nu)}\,r(\tau)}\, \nn
& \times Y_{\ell m}^\ast\!\left(\theta(\tau),\phi(\tau)\right)\,
e^{\,i\Phi(\tau,\nu)}, \label{4.6}
\end{align}
with phase
\begin{equation}
\Phi(\tau,\nu)\equiv \omega \tau + \nu\,t(\tau) + k_\infty(\nu)\,r_*(\tau). \label{4.7}
\end{equation}
Using the geodesic relations \eqref{2.5}--\eqref{2.8} and the near-horizon WKB replacement \(k_\infty(\nu)\to \nu\) inside the rapidly varying part, we may equivalently write
\begin{align}
\Phi(\tau,\nu)&= \omega \tau + \nu\,\left[\,t(\tau)-r_*(\tau)\,\right] + \Delta(\nu)\,r_*(\tau),\nn \Delta(\nu)&\equiv \nu-k_\infty(\nu), \label{4.8}
\end{align}
where \(\Delta(\nu)\) is smooth and slowly varying. The polarization contraction can be organized with the projector defined in \eqref{3.17}. Summing over the three Proca polarizations gives the smooth weight
\begin{align}
\sum_{\lambda=1}^{3}\big|u^\mu\varepsilon^{(\lambda)}_{\mu}\big|^{2}
&= u^\mu u^\nu
\left(-\eta_{\mu\nu}+\frac{k_\mu k_\nu}{m_V^{2}}\right) \nn
&= -\,u^\mu u_\mu+\frac{(u\!\cdot\!k)^{2}}{m_V^{2}}
= -1+\frac{\Omega^{2}(\tau,\nu)}{m_V^{2}}, \label{4.9}
\end{align}
where \(\Omega(\tau,\nu)=-k_\mu u^\mu\) is the comoving frequency along the worldline, equal to the redshift-Doppler factor given in \eqref{2.15} for the near-horizon outgoing branch.

The single-mode excitation probability for the charged-monopole detector then reads
\begin{align}
P_{\text{exc}}^{\text{(mono)}}(\nu,\ell)
&=\sum_{m,\lambda}\Big|\mathcal{A}^{\text{(mono)}}_{\nu\ell m\lambda}\Big|^{2} \nn
&= g_V^{2}\,|m_{eg}|^{2}\,
\Bigg|\int d\tau\,\frac{e^{\,i\Phi(\tau,\nu)}}{\sqrt{k_\infty(\nu)}\,r(\tau)}\,
\mathcal{S}_{\ell m}(\tau,\nu)\Bigg|^{2}, \label{4.10}
\end{align}
where the smooth envelope
\begin{equation}
\mathcal{S}_{\ell m}(\tau,\nu)\equiv
Y_{\ell m}^\ast\!\left(\theta(\tau),\phi(\tau)\right)\,
\left[\,-1+\frac{\Omega^{2}(\tau,\nu)}{m_V^{2}}\,\right]^{1/2} \label{4.11}
\end{equation}
collects the angular dependence and the polarization weight. In the large-\(\omega\) regime and with the cavity isolating a single outgoing channel, the integral in \eqref{4.10} will be evaluated by stationary phase around the near-horizon segment where \(\Phi'(\tau,\nu)=0\), with the rapidly varying part governed by \(-\nu[t(\tau)-r_*(\tau)]\) and the smooth prefactor supplied by \(\mathcal{S}_{\ell m}\) and \(1/r(\tau)\). The appearance of \(\Omega(\tau,\nu)\) in \eqref{4.11} traces directly to the current coupling \(u\!\cdot\!a\) and encodes the distinctive mass/polarization dependence of the vector response.

\subsection{Physical dipole coupling} \label{ssec4.2}
We model a neutral atom with an electric dipole operator orthogonal to its four-velocity. The interaction along the worldline is
\begin{equation}
L_{\text{int}}(\tau) \;=\; -\, d^\mu(\tau)\,F_{\mu\nu}\!\left(x(\tau)\right)\,u^\nu(\tau), \qquad u_\mu d^\mu(\tau)=0, \label{4.12}
\end{equation}
with \(F_{\mu\nu}=\nabla_\mu A_\nu-\nabla_\nu A_\mu\). In the interaction picture,
\begin{align}
&d^\mu(\tau)= d_{eg}\,e^{+i\omega\tau}\,\hat d^\mu\,\sigma_{eg}+ d_{ge}\,e^{-i\omega\tau}\,\hat d^\mu\,\sigma_{ge},\nn &\hat d^\mu \hat d_\mu=-1,\quad u_\mu \hat d^\mu=0. \label{4.13}
\end{align}
Using the properly normalized outgoing modes from Sec. \ref{sec3}, the field strength associated with a single mode \((\nu,\ell,m,\lambda)\) is
\begin{equation}
F_{\mu\nu}^{(\lambda)}(x)\;=\; i\left[k_\mu \,\mathcal{A}^{(\lambda)}_{\nu;\nu\ell m}(x)-k_\nu \,\mathcal{A}^{(\lambda)}_{\mu;\nu\ell m}(x)\right], \label{4.14}
\end{equation}
where \(\mathcal{A}^{(\lambda)}_{\mu;\nu\ell m}\) has the asymptotics in \eqref{4.4} and \(k^\mu=(\nu,\mathbf{k})\) with \(k_\infty=\sqrt{\nu^2-m_V^2}\). Contracting with \(u^\nu\) defines the electric field in the atom's instantaneous rest frame \cite{scully1997quantum,cohen1998atom},
\begin{equation}
E_\mu \;\equiv\; F_{\mu\nu}\,u^\nu \;=\; i\left[-(u\!\cdot\!k)\,\mathcal{A}_\mu + (u\!\cdot\!\mathcal{A})\,k_\mu\right]. \label{4.15}
\end{equation}
For the Proca modes we have \(k\!\cdot\!\varepsilon^{(\lambda)}=0\) (the momentum-space form of \(\nabla_\mu A^\mu=0)\), so \((u\!\cdot\!\mathcal{A})\,k_\mu\) drops out and
\begin{equation}
    E_\mu = i\frac{Y_{\ell m}}{r\sqrt{k_\infty(\nu)}}
     \Big[k_\mu(\varepsilon\cdot u)-\Omega(\tau,\nu)\varepsilon_\mu\Big]
     e^{-i\nu t}e^{-ik_\infty r_*}+\cdots,
     \label{4.16}
\end{equation}
with $\Omega=-k\cdot u$. The counterrotating channel \(|g;0_B\rangle\!\to\!|e;1_{\nu\ell m\lambda}\rangle\) gives the first-order amplitude
\begin{equation}
\mathcal{A}^{\text{(dip)}}_{\nu\ell m\lambda}
= -\,i g_d d_{eg}\!\int d\tau\;
\hat d^\mu\,E_\mu^{(\lambda)\,*}\!\left(x(\tau)\right)\,e^{+i\omega\tau}, \label{4.17}
\end{equation}
which, after inserting \eqref{4.16} and using the cavity-selected outgoing branch, reduces to
\begin{align}
    \mathcal{A}^{\text{(dip)}}_{\nu\ell m\lambda} &= -g_d d_{eg}\int d\tau \frac{e^{i\Phi(\tau,\nu)}}{\sqrt{k_\infty(\nu)}r(\tau)} \nn
    &\times \left[\Omega(\tau,\nu)\hat d\cdot \varepsilon^{(\lambda)}-(\hat d\cdot k)(\varepsilon^{(\lambda)}\cdot u)\right] Y_{\ell m}^\ast(\theta(\tau),\phi(\tau)), \label{4.18}
\end{align}
with the same phase used in \eqref{4.7}--\eqref{4.8},
\begin{align}
\Phi(\tau,\nu)&= \omega \tau + \nu\,t(\tau) + k_\infty(\nu)\,r_*(\tau) \nn
&= \omega \tau + \nu\left[t(\tau)-r_*(\tau)\right]+\Delta(\nu)\,r_*(\tau), \nn \Delta(\nu)&=\nu-k_\infty(\nu). \label{4.19}
\end{align}
Summing over the three physical polarizations can be organized with the projector \(\Pi_{\mu\nu}=-\eta_{\mu\nu}+k_\mu k_\nu/m_V^2\) (cf. \eqref{3.17}),
\begin{equation}
\sum_{\lambda=1}^3 \big|\hat d^\mu \varepsilon^{(\lambda)}_{\mu}\big|^2
=\hat d^\mu \hat d^\nu \,\Pi_{\mu\nu}(\nu,m_V), \label{4.20}
\end{equation}
and, if desired, one can isotropically average \(\hat d^\mu\) in the atom's rest frame to obtain a smooth, order-unity function of \(\nu\). The single-mode excitation probability is therefore
\begin{align}
P_{\text{exc}}^{\text{(dip)}}(\nu,\ell)
&=\sum_{m,\lambda}\Big|\mathcal{A}^{\text{(dip)}}_{\nu\ell m\lambda}\Big|^{2}
= g_d^{2}\,|d_{eg}|^{2}\, \nn
& \times \Bigg|\int d\tau\,\frac{\Omega(\tau,\nu)\,e^{\,i\Phi(\tau,\nu)}}{\sqrt{k_\infty(\nu)}\,r(\tau)}\,
\mathcal{D}_{\ell m}(\tau,\nu)\Bigg|^{2}, \label{4.21}
\end{align}
with the smooth envelope
\begin{equation}
\mathcal{D}_{\ell m}(\tau,\nu)\equiv
Y_{\ell m}^\ast\!\left(\theta(\tau),\phi(\tau)\right)\,
\left[\hat d^\mu \hat d^\nu \,\Pi_{\mu\nu}(\nu,m_V)\right]^{1/2}. \label{4.22}
\end{equation}
Relative to the current coupling in \eqref{4.10}--\eqref{4.11}, the dipole response carries an extra factor \(\Omega(\tau,\nu)\) arising from the electric field \(E_\mu=F_{\mu\nu}u^\nu\). In the large-\(\omega\) regime, the stationary-phase point lies on the near-horizon segment where the rapidly varying phase is controlled by \(-\nu[t(\tau)-r_*(\tau)]\), guaranteeing the same Planckian detailed-balance factor as in the scalar derivation, while the \(\Omega(\tau,\nu)\) and projector in \eqref{4.22} modify only the smooth prefactor and the threshold behavior inherited from \(k_\infty(\nu)\). 

\subsection{Counterrotating channel and selection rules} \label{ssec4.3}
With the Boulware vacuum as the in-state and atoms initially in \(|g\rangle\), only the counterrotating processes (atom excitation accompanied by field emission) survive at first order. The rotating processes either (i) excite the atom while annihilating a quantum \((a_{\nu\ell m\lambda})\), which vanishes on \(|0_B\rangle\), or (ii) deexcite the atom while creating a quantum-irrelevant for atoms prepared in \(|g\rangle\). This is the same mechanism emphasized in Scully \textit{et al.} \cite{Scully:2017utk} for the scalar case and is the operative origin of emission-before-absorption. 

For the current coupling \eqref{4.1} the relevant vertex on \(|g;0_B\rangle\) is
\begin{equation}
\mathcal{V}^{\text{(mono)}}_{\rm cr}(\tau)\;=\;g_V\,m_{eg}\,e^{+i\omega\tau}\,u^\mu(\tau)\,A_\mu^{(+)\,\dagger}\!\left(x(\tau)\right), \label{4.23}
\end{equation}
and for the dipole coupling \eqref{4.12}
\begin{equation}
\mathcal{V}^{\text{(dip)}}_{\rm cr}(\tau)\;=\;g_d\,d_{eg}\,e^{+i\omega\tau}\,\hat d^\mu(\tau)\,F_{\mu\nu}^{(+)\,\dagger}\!\left(x(\tau)\right)\,u^\nu(\tau), \label{4.24}
\end{equation}
where the \((+)\) parts denote the positive-Killing-frequency mode operators. The corresponding amplitudes have already been reduced to the single worldline integrals in \eqref{4.6} and \eqref{4.18}; their phases are governed by \(-\nu[t(\tau)-r_*(\tau)]\), which is what ultimately produces the Planckian detailed-balance factor as in the scalar derivation. 

The physically relevant selection rules are:
\begin{enumerate}
    \item Initial \(|g;0_B\rangle\) forbids the rotating \(\propto a_{\nu\ell m\lambda}\sigma_{eg}\) channel; only the counterrotating \(\propto a^\dagger_{\nu\ell m\lambda}\sigma_{eg}\) contributes to excitation. This singles out the terms displayed in \eqref{4.23}--\eqref{4.24} and underlies the emission-before-absorption picture adopted in the master-equation treatment.
    \item Asymptotically, propagation requires
    \begin{equation}
        \nu\ge m_V,\qquad k_\infty(\nu)=\sqrt{\nu^2-m_V^2}, \label{4.25}
   \end{equation}
   so modes below threshold do not contribute to the escaping flux. In the stationary-phase evaluation, the rapidly varying piece depends only on \(\nu>0\), thus preserving the same detailed-balance structure as in Eqs. (B13)-(B15) of \cite{Scully:2017utk}.
   \item The coupling to a given \((\ell,m)\) is weighted by \(Y_{\ell m}(\theta(\tau),\phi(\tau))\). For a narrow, axially aligned, radially infalling beam (as in the pencil-like cloud of Fig. 1), the trajectory satisfies \(\phi=\) const and \(\theta\approx 0\), which selects predominantly \(m=0\) and suppresses \(|m|\ge1\) through the small-angle behavior of \(Y_{\ell m}\). The cavity's mode selector then isolates a single \((\ell, m)\) channel, exactly as in the scalar analysis.
   \item For the monopole coupling, the polarization sum produces
   \begin{equation}
    \sum_{\lambda}\!\big|u\!\cdot\!\varepsilon^{(\lambda)}\big|^2
   =-1+\frac{\Omega^2(\tau,\nu)}{m_V^2}, \label{4.26}
   \end{equation}
   where \(\Omega=-k\!\cdot\!u\) grows without bound near the horizon; this enhances the longitudinal contribution of the massive vector relative to the massless limit and feeds only a smooth prefactor in the final probability. For the dipole coupling the contraction \(\hat d^\mu \hat d^\nu \Pi_{\mu\nu}\) projects onto the spatially transverse subspace in the atom's rest frame, favoring the two transverse polarizations and weighting them by the dipole orientation distribution. In both cases, axial/polar sectors contribute with different greybody barriers, but the near-horizon phase (and hence the thermal factor) remains spin-independent.
   \item Since all master fields reduce to plane waves \(\sim e^{-i\nu(t-r_*)}\) as \(r\to r_g^+\), the same stationary-phase structure that yielded Eq. (B14) in \cite{Scully:2017utk} in the scalar case recurs here, while the mass and polarization affect only the prefactor (and the \(\nu\ge m_V\) threshold).
\end{enumerate}
These rules ensure that, after summing over \((m,\lambda)\) and implementing the single-mode cavity filter, the excitation probabilities reduce to the forms already prepared in \eqref{4.10}--\eqref{4.11} and \eqref{4.21}--\eqref{4.22}. When we convert these amplitudes into rates and feed them into the master equation, we will recover the same detailed-balance ratio \(\Gamma_e/\Gamma_a=e^{-2\xi}\) with \(\xi=2\pi r_g\nu/c\), exactly mirroring the scalar result while carrying the Proca-specific prefactors and the hard mass threshold.

With the emission amplitudes written using \(+\nu\left[t(\tau)-r_*(\tau)\right]\) in the rapidly varying phase, the stationary-phase evaluation repeats Scully \textit{et al.}'s steps [their (B11)–(B14)] verbatim \cite{Scully:2017utk}, yielding the same Planckian factor, while the Proca mass/polarization affects only the smooth prefactor and the \(\nu\ge m_V\) threshold.

\section{Excitation probability along the infall worldline} \label{sec5}
We evaluate the first-order excitation amplitudes derived in Sec. \ref{sec4} along the infalling geodesic. For both detector models, the integrals are dominated by the near-horizon segment where the phase is governed by the combination \(t-r_*\). Using the geodesic relations from Sec. \ref{sec2} and the unit-flux mode normalization from Sec. \ref{sec3}, the worldline integral reduces, in the large-\(\omega\) regime, to an end point (near-horizon) asymptotics that yields a gamma-function. Its modulus square produces the Planckian detailed-balance factor \(\left(e^{4\pi \nu}-1\right)^{-1}\). Mass and polarization enter only through smooth prefactors and the asymptotic kinematic threshold \(\nu\ge m_V\).

\subsection{General stationary-phase integral} \label{ssec5.1}
For a single outgoing mode \((\nu,\ell,m,\lambda)\), the first-order counterrotating amplitude has the structure
\begin{equation}
\mathcal{A}_\bullet(\nu,\ell,m,\lambda)
=\int_{-\infty}^{+\infty}\! d\tau\;\mathcal{G}_\bullet(\tau,\nu,\ell,m,\lambda)\,
e^{\,i\Phi(\tau,\nu)}, \label{5.1}
\end{equation}
where \(\bullet\in{\text{mono},\text{dip}}\) denotes the detector model. The smooth envelope \(\mathcal{G}_\bullet\) collects the \(1/[r(\tau)\sqrt{k_\infty(\nu)}]\) factor from mode normalization and the appropriate polarization weight (cf. \eqref{4.11} for the current coupling and its dipole analogue), while the phase is
\begin{equation}
\Phi(\tau,\nu)=\omega\tau+\nu\,\left[t(\tau)-r_*(\tau)\right]+\text{(constant)} \label{5.2}
\end{equation}
We now isolate the rapidly varying near-horizon piece. Let \(\tau_h\) be the finite proper time at which the geodesic reaches \(r=1\). Set \(y\equiv \tau_h-\tau>0\) and use the geodesic relations
\begin{equation}
\frac{dr}{d\tau}=-\frac{1}{\sqrt r},\quad
\frac{dt}{d\tau}=\frac{r}{r-1},\quad
\frac{dr_*}{d\tau}=\frac{dr}{d\tau}\,\frac{dr_*}{dr}=-\frac{\sqrt r}{\,r-1\,}. \label{5.3}
\end{equation}

Near the horizon, let $r(\tau) = 1+\delta$ with $0 < \delta \ll 1$. From the radial geodesic equation $dr/d\tau = -1/\sqrt{r}$, we see that $dr/d\tau \to -1$ as $r \to 1$, which implies that the coordinate distance from the horizon scales linearly with proper time: $\delta \approx \tau_h - \tau \equiv y$. 
We now expand the phase derivative using the redshift factor:
\begin{equation}
\frac{d}{d\tau}(t-r_*) = \frac{r+\sqrt{r}}{r-1} \approx \frac{1+(1+\delta/2)}{\delta} \approx \frac{2}{y}. \label{5.3a}
\end{equation}
Integrating this expression with respect to $\tau$ (noting that $d\tau = -dy$) yields the universal logarithmic mapping characteristic of the horizon:
\begin{equation}
t(\tau)-r_*(\tau)= -\,2\ln y + C_h + \mathcal O(y), \qquad y\to 0^{+}, \label{5.4}
\end{equation}
consistent with surface gravity \(\kappa=\frac{1}{2}\) in our dimensionless units. Substituting \eqref{5.4} into \eqref{5.2} and absorbing constants into an overall phase,
\begin{equation}
\Phi(\tau,\nu)= \Phi_h - \omega y - 2\nu\ln y + \Delta(\nu)\,r_*(\tau_h) + \mathcal O(y). \label{5.5}
\end{equation}
Since \(\mathcal{G}_\bullet(\tau,\nu,\ell,m,\lambda)\) is smooth, we replace it by its horizon value \(\mathcal{G}_\bullet(\tau_h;\nu,\ell,m,\lambda)\equiv \mathcal{G}_{\bullet h}(\nu,\ell,m,\lambda)\) at leading order. The amplitude \eqref{5.1} becomes
\begin{equation}
\mathcal{A}_\bullet(\nu,\ell,m,\lambda)=\mathcal{N}_\bullet(\nu,\ell,m,\lambda)\,
\underbrace{\int_{0}^{\infty}\! dy\; y^{-2i\nu}\,e^{-i\omega y}}_{\displaystyle \mathcal{I}(\nu,\omega)}, \label{5.6}
\end{equation}
where \(\mathcal{N}_\bullet\) is a smooth prefactor containing the polarization weight at the horizon, the \(1/\sqrt{k_\infty(\nu)}\) factor, and the harmless phase \(e^{i[\Phi_h+\Delta(\nu) r_*(\tau_h)]}\). We note that this integral is dominated by the logarithmic phase singularity at the horizon ($y=0$), rather than a conventional stationary point on the real axis.

The elementary Laplace-type integral in \eqref{5.6} is expressed in terms of the gamma function,
\begin{equation}
I(\nu,\omega)=(i\omega)^{-(1-2i\nu)}\Gamma(1-2i\nu), \qquad \omega>0. \label{5.7}
\end{equation}
The excitation probability for the selected $(\nu,\ell)$ channel (after summing over $m,\lambda$) therefore contains the universal factor
\begin{align}
|\mathcal{I}(\nu,\omega)|^{2}&=\frac{1}{\omega^2} \left| \Gamma(1-2i\nu) \right|^2
=\frac{2\pi\nu}{\omega^{2}}\;\frac{1}{\sinh(2\pi\nu)} \nn
&=\frac{4\pi\nu}{\omega^{2}}\;\frac{1}{e^{2\pi\nu}-e^{-2\pi\nu}}\;. \label{5.8}
\end{align}
Equation \eqref{5.8} governs the thermal character of the problem. While $|\mathcal{I}|^2$ itself follows a statistics-dependent profile (here proportional to $\text{cosech}(2\pi\nu)$), the crucial feature for thermodynamics is the relation between excitation (emission) and de-excitation (absorption). 
All dependence on the detector model (current vs. dipole), the Proca mass and polarizations, and the asymptotic threshold $\nu\ge m_V$ resides in the smooth prefactor $\mathcal{N}_\bullet$ and in $k_\infty(\nu)$. 
Consequently, at leading order in $1/\omega$, the ratio of probabilities satisfies the detailed balance condition $P_{\text{exc}}/P_{\text{abs}} = e^{-4\pi\nu}$ (derived below via the KMS condition), ensuring the steady state is thermal.

\subsection{Universality of the thermal factor} \label{ssec5.2}
The Planckian denominator obtained in \eqref{5.8} is independent of the detector model (current vs. dipole), the field's spin, the Proca mass, and greybody details. The reason is structural: after isolating a single outgoing Schwarzschild mode, every first-order counterrotating amplitude reduces to an integral of the form
\begin{equation}
\int d\tau\,\underbrace{\mathcal{G}(\tau,\nu)}_{\text{smooth}}\,
\exp\!\big\{i\omega\tau+i\nu\,\left[t(\tau)-r_*(\tau)\right] + i\Delta(\nu)r_*(\tau)\Big\}, \label{5.9}
\end{equation}
where \(\mathcal{G}\) is a smooth, slowly varying function that encodes the detector-specific contraction (projectors, \(\Omega=-k\!\cdot\!u\), orientation factors), the \(1/\sqrt{k_\infty(\nu)}\) from unit-flux normalization, and any cavity-induced mode weights. None of these ingredients alters the phase driver \(t-r_*\).

Along the infall worldline, the near-horizon mapping of proper time to Killing time/tortoise time is universal:
\begin{equation}
t(\tau)-r_*(\tau)=-\,2\ln(\tau_h-\tau)+C_h+\mathcal O(\tau_h-\tau), \label{5.10}
\end{equation}
with surface gravity \(\kappa=\frac{1}{2}\) in our dimensionless units. Introducing \(y=\tau_h-\tau>0\) and freezing \(\mathcal{G}\) at its horizon value produces the Laplace-type integral
\begin{equation}
\int_0^\infty dy\, y^{-2i\nu}\,e^{-i\omega y}, \label{5.11}
\end{equation}
whose exact evaluation is \(\mathcal{I}(\nu,\omega)= e^{-i\pi/2}\,\omega^{-1+2i\nu}\Gamma(1-2i\nu)\)\, yielding
\begin{equation}
\;|\mathcal{I}(\nu,\omega)|^2=\frac{4\pi\nu}{\omega^2}\,\frac{1}{e^{4\pi\nu}-1}\;. \label{5.12}
\end{equation}
Equation \eqref{5.12} is the universal thermal factor for the excitation probabilities. All dependence on the detector realization (current vs. dipole), polarization content, Proca mass and threshold \(\nu\ge m_V\), and barrier transmission enters only through the smooth prefactor that multiplies \eqref{5.12}.

\begin{figure}
    \centering
    \includegraphics[width=\columnwidth]{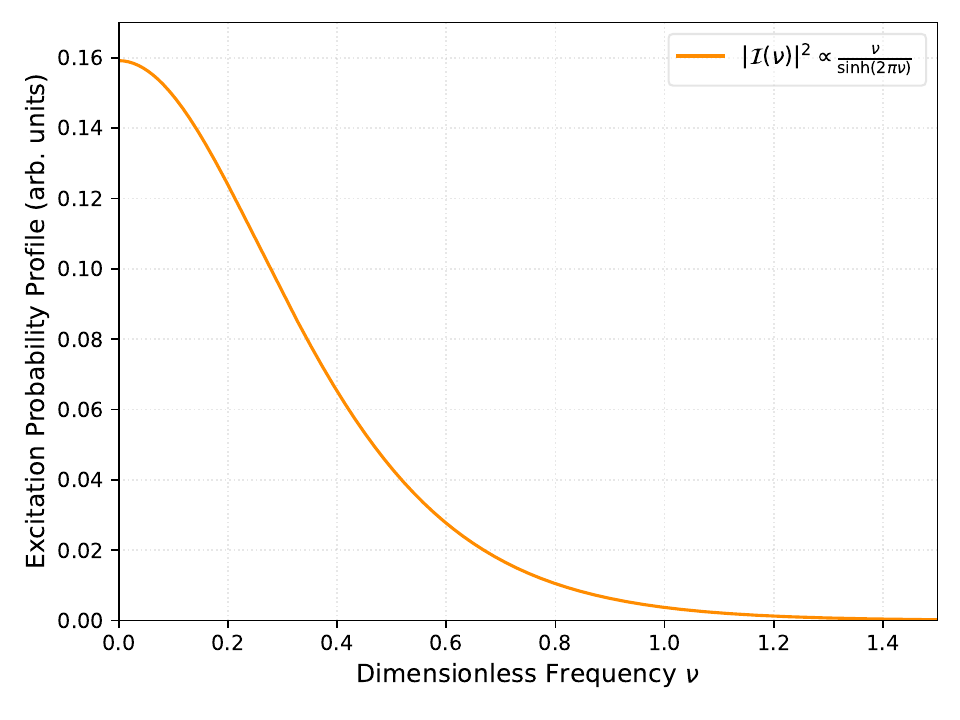}
    \caption{The universal thermal kernel $|\mathcal{I}(\nu)|^2$ as a function of dimensionless frequency $\nu$. The profile exhibits a cosech-like decay characteristic of the near-horizon Rindler mapping, independent of the field's spin or mass.}
    \label{fig2}
\end{figure}

The universality of the excitation mechanism is displayed in Fig. \ref{fig2}, which plots the squared modulus of the near-horizon integral $|\mathcal{I}(\nu)|^2$. This profile, governed by the $\text{cosech}(2\pi\nu)$ dependence derived in Eq. \eqref{5.8}, dictates the thermal response of the detector. Crucially, this curve is identical for both the monopole and dipole couplings and for any mass $m_V$, confirming that the thermal character arises solely from the geometric analyticity of the Rindler time coordinate $t-r_*$ rather than the specific details of the probe or the field. The detailed balance ratio required for thermal equilibrium emerges from the symmetry of this kernel under $\nu \to -\nu$.

The same near-horizon analyticity that leads to \eqref{5.12} also fixes detailed balance between the counterrotating excitation channel and its absorption counterpart (appearing later in the master equation). Under \(\nu\to -\nu\) the logarithmic phase in \eqref{5.10} implements the (Kubo-Martin-Schwinger) KMS-type periodicity of Rindler time, giving
\begin{equation}
\;P_{\rm abs}(\nu)=e^{\,4\pi\nu}\,P_{\rm exc}(\nu)\;, \label{5.13}
\end{equation}
with \(\nu>0\). Thus the ratio is universal, while \(P_{\rm exc}\) itself carries the model-dependent prefactor and the Proca threshold via \(k_\infty(\nu)=\sqrt{\nu^2-m_V^2}\).

\subsection{Prefactors, thresholds, and detailed balance} \label{ssec5.3}
The near-horizon integral in \eqref{5.6}--\eqref{5.8} fixes the entire \(\omega\)- and \(\nu\)-dependence that carries the thermal factor. Everything else, from detector realization, polarization structure, greybody transmission, and the Proca mass, enters multiplicatively through a smooth prefactor evaluated on the horizon segment, together with the asymptotic kinematic threshold \(\nu\ge m_V\).

For a single, cavity-selected outgoing channel \((\nu,\ell)\) we write, after summing over \(m\) and polarizations and retaining the leading term in the large-\(\omega\) expansion,
\begin{align}
P_{\text{exc}}^{\bullet}(\nu,\ell)
&=\frac{4\pi\nu}{\omega^{2}}\;
\frac{\mathcal{K}_\bullet\,\Xi_{\ell}(\nu)}{e^{4\pi\nu}-1}\; \nn
& \times \frac{\Theta(\nu-m_V)}{k_\infty(\nu)}\;
\overline{\mathcal{P}}_{\bullet}(\nu,\ell)
\left[1+\mathcal{O}\!\left(\tfrac{\nu}{\omega}\right)\right], \nn
k_\infty(\nu)&=\sqrt{\nu^2-m_V^2}, \label{5.14}
\end{align}
where \(\mathcal{K}_\bullet\) collects the detector coupling and matrix element \((\mathcal{K}_{\text{mono}}=g_V^2|m_{eg}|^2\), \(\mathcal{K}_{\text{dip}}=g_d^2|d_{eg}|^2)\), \(\Xi_{\ell}(\nu)\) is the single-channel transmission (greybody) through the exterior potential barrier, \(k_\infty(\nu)^{-1}\) comes from the unit-flux mode normalization at infinity, and \(\overline{\mathcal{P}}_{\bullet}(\nu,\ell)\) is a smooth, order-unity polarization/geometry factor fixed by the corresponding contraction in Sec. \ref{sec4} and by the cavity's angular selection.

Specializing \eqref{5.14} to the two detector models requires handling the sum over $m$. For an atom falling radially along the symmetry axis ($\theta \approx 0$), the interaction is weighted by $Y_{\ell m}(\theta \approx 0)$, which vanishes for $m \neq 0$. Consequently, the sum over $m$ collapses to the single $m=0$ term. With this selection, the probabilities become
\begin{align}
P_{\text{exc}}^{\text{(mono)}}(\nu,\ell)
&=\frac{4\pi\nu}{\omega^{2}}\;
\frac{g_V^{2}|m_{eg}|^{2}}{e^{4\pi\nu}-1}\;
\frac{\Xi_{\ell}(\nu)}{k_\infty(\nu)}\; \nn
& \times \Theta(\nu-m_V)\;
\overline{\mathcal{P}}_{\text{mono}}(\nu,\ell,m=0)
\;, \label{5.15}
\end{align}
\begin{align}
P_{\text{exc}}^{\text{(dip)}}(\nu,\ell)
&=\frac{4\pi\nu}{\omega^{2}}\;
\frac{g_d^{2}|d_{eg}|^{2}}{e^{4\pi\nu}-1}\;
\frac{\Xi_{\ell}(\nu)}{k_\infty(\nu)}\; \\
& \times \Theta(\nu-m_V)\;
\overline{\mathcal{P}}_{\text{dip}}(\nu,\ell,m=0)
\;.
\label{5.16}
\end{align}
The factors \(\overline{\mathcal{P}}_{\text{mono}}\,\overline{\mathcal{P}}_{\text{dip}}\) encode, respectively, the current contraction and the electric-field/dipole contraction at the horizon, including the polarization projector and any orientation averages. Their precise form is not needed to get the thermal law; they modify only the overall spectral weight and are smooth functions of \(\nu\).

The hard mass threshold is enforced by \(\Theta(\nu-m_V)\); close to threshold the asymptotic momentum \(k_\infty(\nu)\) suppresses the probability as \(k_\infty^{-1}\), while the greybody factor \(\Xi_{\ell}(\nu)\) further damps low \(\nu\). In the massless limit \(m_V\to0\) one has \(k_\infty\to\nu\) so the prefactor tends to the familiar \(1/\nu\) normalization, reproducing the scalarlike spectral scaling at fixed \(\Xi_{\ell}(\nu)\) and \(\overline{\mathcal{P}}_{\bullet}\).

Detailed balance follows from the same near-horizon analyticity that produced \eqref{5.12}. Replacing \(\nu\to -\nu\) in the phase reverses the sign of the logarithm and yields
\begin{equation}
P_{\text{abs}}^{\bullet}(\nu,\ell)= e^{\,4\pi\nu}\,P_{\text{exc}}^{\bullet}(\nu,\ell), \qquad \nu>0, \label{5.17}
\end{equation}
so the ratio is universal,
\begin{equation}
\;\frac{P_{\text{exc}}^{\bullet}(\nu,\ell)}{P_{\text{abs}}^{\bullet}(\nu,\ell)}
=\frac{1}{e^{\,4\pi\nu}}\;. \label{5.18}
\end{equation}
Equations \eqref{5.15}--\eqref{5.18} are the inputs we will promote to rates in the next section by dividing by the interaction time set by the cavity transit and by incorporating the single-mode selection, thereby closing the loop to the master-equation description.

\section{Cavity, rates, and master equation} \label{sec6}
We now promote the single-pass excitation probabilities into rates by (i) specifying how the cavity selects one outgoing Schwarzschild mode and how much of it escapes to infinity through the exterior potential (greybody factor), and (ii) dividing by the relevant interaction time associated with the atom's transit across the mode region. The Proca field introduces polarization-dependent greybody transmission: the axial channel is single, while the polar sector yields two effective channels after local diagonalization. Near the horizon all channels share the same plane wave phase responsible for the thermal factor; the greybody matrices merely supply smooth, frequency-dependent multipliers in the prefactor and impose the mass threshold \(\nu\ge m_V\). These ingredients feed directly into the emission/absorption rates used in the master equation.

\subsection{Mode selection and greybody factors} \label{ssec6.1}
We confine the field between tortoise radii \(r_*^{(b)}<r_*<r_*^{(t)}\) with perfect mirrors, then take the standard limits
\begin{equation}
r_b\to 1^{+},\qquad r_t\to \infty, \label{6.1}
\end{equation}
so that the lower mirror approaches the horizon and the upper mirror recedes, leaving a single, cavity-selected outgoing channel. In this limit the standing-wave quantization \(k_\infty L_*=\pi n\) becomes a continuum with density \(\rho(\nu)=\tfrac{L_*}{\pi}\tfrac{\nu}{k_\infty}\,\Theta(\nu-m_V)\). The cavity enforces mode purity (one \((\nu,\ell)\) channel), while the exterior curvature supplies a barrier that partially reflects the wave before it reaches \(\mathscr{I}^+\).

For each \(\ell\) the master radial equations of Sec. \ref{sec3} define an elastic scattering problem on the line of \(r_*\). Let \(\Psi^{\text{in}}_{\ell}\) denote a unit-flux wave incident from the near-horizon side. The Wronskian constancy fixes the scattering matrix and the transmission coefficient \( \mathcal T_\ell(\nu)\in[0,1]\). For the Proca field, the axial sector provides one channel with transmission \( \mathcal T_\ell^{\text{(ax)}}(\nu)\) for \(\ell\ge1\). On the other hand, the polar sector yields two effective channels after local diagonalization with transmissions \( \mathcal T_\ell^{\text{(pol)}\pm}(\nu)\) (for \(\ell\ge0\), noting that \(\ell=0\) is polar only).

{\color{black}At this stage it is important to distinguish inclusion from explicit evaluation. The present paper does not solve the Proca transmission problem numerically for each \((\ell,\nu)\); rather, it formulates the escaping HBAR flux in terms of the exact transmission coefficients associated with the Schwarzschild radial scattering problem. This is sufficient for the purposes of the present analysis because the universal thermal factor was already fixed in Sec. \ref{sec5} by the near-horizon worldline integral, whereas the exterior barrier enters only through the smooth escape factor defined below.} The single-mode escape factor multiplying probabilities is then
\begin{equation}
\Xi_\ell(\nu)=
\begin{cases}
\mathcal T_\ell^{\text{(ax)}}(\nu), & \text{axial channel},\\
\mathcal T_\ell^{\text{(pol)}+}(\nu)+\mathcal T_\ell^{\text{(pol)}-}(\nu), & \text{polar sector}.
\end{cases} \label{6.2}
\end{equation}
Near the horizon, the axial/polar distinction is immaterial for the thermal factor; it reenters only through \(\Xi_\ell(\nu)\) as a smooth function of \(\nu\). {\color{black}Accordingly, the present work establishes how polarization-resolved greybody transmission modifies the HBAR prefactor and escaping flux in full generality, while leaving a dedicated numerical evaluation of the axial/polar transmission functions for future study.}

Two universal limits will be repeatedly used
\begin{equation}
\Xi_\ell(\nu)\rightarrow[\nu\to m_V^{+}]{} 0, \qquad
\Xi_\ell(\nu)\rightarrow[\nu\gg \max{m_V,\ell/r_g}]{} 1. \label{6.3}
\end{equation}
The threshold behavior follows from \(k_\infty(\nu)=\sqrt{\nu^2-m_V^2}\to0^{+}\), which suppresses tunneling through the centrifugal/mass barrier. In practice, close to threshold one finds a power-law suppression governed by the lowest available \(\ell\) in each sector; for example, the absence of \(\ell=0\) in the axial sector implies a stronger low-\(\nu\) suppression there than in the polar sector, where \(\ell=0\) is allowed.

Because our mode functions are normalized to unit outgoing flux at infinity, the replacement probability \(\to\) escaping probability is accomplished by multiplying the near-horizon stationary-phase result by \(\Xi_\ell(\nu)\). Concretely, with the universal integral \(|\mathcal I|^{2}=\tfrac{4\pi\nu}{\omega^{2}},(e^{4\pi\nu}-1)^{-1}\) and the unit-flux normalization that produced the \(k_\infty(\nu)^{-1}\) factor, the cavity-selected, escaping-mode excitation probabilities take the schematic form
\begin{equation}
P_{\text{exc,esc}}^{\bullet}(\nu,\ell)
=\frac{4\pi\,\nu}{\omega^{2}}\,
\frac{\mathcal K_\bullet}{e^{4\pi\nu}-1}\,
\frac{\Theta(\nu-m_V)}{k_\infty(\nu)}\,
\underbrace{\Xi_\ell(\nu)\,\overline{\mathcal P}_\bullet(\nu,\ell)}_{\text{smooth prefactor}}, \label{6.4}
\end{equation}
where \(\mathcal K_\bullet\) is the detector coupling factor and \(\overline{\mathcal P}_\bullet\) is the polarization/geometry weight from Sec. \ref{sec4}. The factor \(\Xi_\ell(\nu)\) will remain attached when we convert probabilities to rates and then to the master equation, ensuring that only the fraction of quanta that traverse the exterior potential contribute to the outgoing flux at \(\mathscr{I}^+\). Noting explicitly, \(\Xi_\ell(\nu)\sim k_\infty^{2\ell+1}\) near threshold generically, so \(\Xi_\ell/k_\infty\sim k_\infty^{2\ell}\) is finite/vanishing (and never divergent).

\subsection{Emission and absorption rates} \label{ssec6.2}
We now convert the single-pass escaping probabilities in \eqref{6.4} into mode rates by multiplying by the atomic throughput across the cavity. Let \(R_{\!a}\) denote the number of atoms (prepared in \(|g\rangle)\) entering the cavity per unit Killing time \(t\) (dimensionless as in Sec. \ref{sec2}). For a cavity-selected outgoing channel \((\nu,\ell)\) the emission rate is
\begin{align}
\;
&\Gamma_{e}^{\bullet}(\nu,\ell)
= R_{\!a}\;P_{\text{exc,esc}}^{\bullet}(\nu,\ell) \nn
&= R_{\!a}\,\frac{4\pi\,\nu}{\omega^{2}}\,
\frac{\mathcal K_\bullet}{e^{4\pi\nu}-1}\,
\frac{\Theta(\nu-m_V)}{k_\infty(\nu)}\,
\Xi_\ell(\nu)\,\overline{\mathcal P}_\bullet(\nu,\ell)\;, \label{6.5}
\end{align}
with \(k_\infty(\nu)=\sqrt{\nu^2-m_V^2}\), \(\mathcal K_{\text{mono}}=g_V^2|m_{eg}|^2\), \(\mathcal K_{\text{dip}}=g_d^2|d_{eg}|^2\), and \(\overline{\mathcal P}_\bullet\) the smooth polarization/geometry factor defined in Sec. \ref{sec4}. The absorption rate follows from the universal detailed balance established in Sec. \ref{sec5},
\begin{equation}
\;\Gamma_{a}^{\bullet}(\nu,\ell)=e^{\,4\pi\nu}\,\Gamma_{e}^{\bullet}(\nu,\ell);\, \label{6.6}
\end{equation}
so the ratio \(\Gamma_{e}^{\bullet}/\Gamma_{a}^{\bullet}=e^{-4\pi\nu}\) is independent of detector realization, mass, or greybody weights \cite{Kubo:1957mj,Martin:1959jp,breuer2002theory}. Note that \(R_a\) affects absolute fluxes and relaxation speeds, but cancels from \(\langle n\rangle_{\mathrm{ss}}\).

For bookkeeping in the master equation it is convenient to absorb the throughput into an effective coupling,
\begin{equation}
\kappa_\bullet \;\equiv\; R_{\!a}\,\mathcal K_\bullet, \label{6.7}
\end{equation}
so that
\begin{align}
\Gamma_{e}^{\bullet}(\nu,\ell)
&= \frac{4\pi\,\nu}{\omega^{2}}\,
\frac{\kappa_\bullet}{e^{4\pi\nu}-1}\,
\frac{\Theta(\nu-m_V)}{k_\infty(\nu)}\,
\Xi_\ell(\nu)\,\overline{\mathcal P}_\bullet(\nu,\ell), \nn
\Gamma_{a}^{\bullet}(\nu,\ell)&=e^{\,4\pi\nu}\,\Gamma_{e}^{\bullet}(\nu,\ell). \label{6.8}
\end{align}
When several \(\ell\)-channels are simultaneously admitted by the cavity filter, we sum the escaping rates,
\begin{equation}
\Gamma_{e,a}^{\bullet}(\nu)=\sum_{\ell}\Gamma_{e,a}^{\bullet}(\nu,\ell), \label{6.9}
\end{equation}
bearing in mind that axial/polar channels contribute through their respective \(\Xi_\ell(\nu)\) and (implicitly) through \(\overline{\mathcal P}_\bullet(\nu,\ell)\). In the massless limit \(m_V\to0\) one has \(k_\infty\to \nu\) and the threshold factor becomes trivial, recovering the familiar \(\propto \nu/\omega^2\) scaling of the scalarlike response (up to the polarization prefactor). These \(\Gamma_{e,a}\) are the inputs for the mode-occupancy master equation in the next subsection.

\subsection{Master equation and steady state} \label{ssec6.3}
For the single, cavity-selected outgoing mode of frequency \(\nu\), we model the coarse-grained field dynamics with a birth-death master equation for the diagonal elements \(\rho_{nn}(t)\equiv\langle n|\rho(t)|n\rangle\) in the Fock basis. With emission and absorption rates \(\Gamma_e^\bullet(\nu,\ell)\) and \(\Gamma_a^\bullet(\nu,\ell)\) from \eqref{6.8}, the population dynamics reads
\begin{align}
\frac{d\rho_{nn}}{dt}
&= -\,\Gamma_e\left[(n+1)\rho_{nn}-n\,\rho_{n-1,n-1}\right] \nn
&-\,\Gamma_a\left[n\,\rho_{nn}-(n+1)\rho_{n+1,n+1}\right], \label{6.10}
\end{align}
where, to declutter notation, we have fixed \((\nu,\ell)\) and the detector label \(\bullet\) and written \(\Gamma_{e,a}\equiv\Gamma_{e,a}^\bullet(\nu,\ell)\). This equation is equivalent to the standard Lindblad form with jump operators \(a^\dagger\) and \(a\) \cite{Lindblad:1975ef},
\begin{equation}
\dot\rho \;=\; \Gamma_e\!\left(a^\dagger \rho a - \frac{1}{2}{aa^\dagger\,\rho}\right)
+ \Gamma_a\!\left(a \rho a^\dagger - \frac{1}{2}{a^\dagger a\,\rho}\right), \label{6.11}
\end{equation}
and it preserves positivity and trace.

The detailed-balance relation proved in \eqref{5.17}--\eqref{5.18} implies \(\Gamma_e/\Gamma_a=e^{-4\pi\nu}\). The steady state \(\dot\rho_{nn}=0\) of \eqref{6.10} is therefore geometric:
\begin{equation}
\frac{\rho_{n+1,n+1}^{(\text{ss})}}{\rho_{nn}^{(\text{ss})}}
=\frac{\Gamma_e}{\Gamma_a}=e^{-4\pi\nu}
\;\Longrightarrow\;
\rho_{nn}^{(\text{ss})}=(1-e^{-4\pi\nu})\,e^{-4\pi\nu\,n}. \label{6.12}
\end{equation}
The mean mode occupancy follows either from \eqref{6.10} or directly from \eqref{6.12}:
\begin{align}
\frac{d\langle n\rangle}{dt}&= -(\Gamma_a-\Gamma_e)\,\langle n\rangle + \Gamma_e, \nn
\;\langle n\rangle_{\text{ss}}&=\frac{\Gamma_e}{\Gamma_a-\Gamma_e}
=\frac{1}{e^{4\pi\nu}-1}\;. \label{6.13}
\end{align}
Crucially, \(\langle n\rangle_{\text{ss}}\) is independent of all prefactors (detector realization, polarization projectors, greybody transmissions, and the Proca mass) because these enter \(\Gamma_{e,a}\) multiplicatively and cancel in the ratio fixed by detailed balance. They do, however, set the approach rate to stationarity via \(\Gamma_a-\Gamma_e\):
\begin{equation}
\langle n(t)\rangle-\langle n\rangle_{\text{ss}}
=\left(\langle n(0)\rangle-\langle n\rangle_{\text{ss}}\right)\,
e^{-(\Gamma_a-\Gamma_e)t}. \label{6.14}
\end{equation}
When several \(\ell\)-channels are admitted, the total master equation is a direct sum of copies of \eqref{6.10}, one per channel, with \(\Gamma_{e,a}(\nu,\ell)\) as in \eqref{6.8}. Each channel relaxes to the same geometric law \eqref{6.12} with the same \(\langle n\rangle_{\text{ss}}\) in \eqref{6.13}, while differing only in relaxation rate and overall flux set by \(\Xi_\ell(\nu)\), \(k_\infty(\nu)\), and the detector-dependent prefactor carried in \(\Gamma_{e,a}\).

\section{HBAR entropy for massive spin-1 quanta} \label{sec7}
We compute the entropy carried away by the escaping Proca quanta selected by the cavity. The master equation fixes both the mode populations and the detailed-balance ratio. Using a near-stationary substitution for the logarithm of the density matrix, we obtain a simple, universal entropy-flux law: the entropy rate for each mode equals \(4\pi k_B\nu\) times the net particle flux in that mode. All Proca-specific physics (mass threshold, polarization sums, greybody transmission, detector realization) enters only through the net flux; the coefficient \(4\pi\nu\) is the same as in the scalar case and follows from the near-horizon phase.

\subsection{Entropy flux from the master equation} \label{ssec7.1}
For a single, cavity-selected outgoing mode of frequency \(\nu\), the diagonal master equation reads
\begin{align}
\frac{d\rho_{nn}}{dt}
&= -\,\Gamma_e\left[(n+1)\rho_{nn}-n\,\rho_{n-1,n-1}\right] \nn
&-\,\Gamma_a\left[n\,\rho_{nn}-(n+1)\rho_{n+1,n+1}\right], \label{7.1}
\end{align}
with emission and absorption rates \(\Gamma_{e}\) and \(\Gamma_{a}\) given in terms of the escaping probabilities (and hence containing the Proca prefactors). The mean occupancy evolves as
\begin{equation}
\frac{d\langle n\rangle}{dt}= -(\Gamma_a-\Gamma_e)\,\langle n\rangle + \Gamma_e. \label{7.2}
\end{equation}
We define the net particle flux in the mode by
\begin{equation}
\dot{\bar n}_\nu \;\equiv\; \frac{d\langle n\rangle}{dt}. \label{7.3}
\end{equation}

The von Neumann entropy of the mode is \(S=-k_B\sum_{n}\rho_{nn}\ln\rho_{nn}\). Its exact time derivative is
\begin{equation}
\frac{dS}{dt}
=-k_B\sum_{n}\dot\rho_{nn}\,\ln\rho_{nn}. \label{7.4}
\end{equation}
Following the standard near-stationary replacement, we evaluate the logarithm at the instantaneous steady state corresponding to the same rates in \eqref{7.1}. The steady state solves \(\Gamma_e/\Gamma_a=e^{-4\pi\nu}\) with
\begin{equation}
\rho_{nn}^{(\mathrm{ss})}=(1-e^{-4\pi\nu})\,e^{-4\pi\nu\,n}. \label{7.5}
\end{equation}
Using \(\ln\rho_{nn}^{(\mathrm{ss})}=\ln(1-e^{-4\pi\nu})-4\pi\nu\,n\) in \eqref{7.4}, the constant term drops by probability conservation \(\sum_n\dot\rho_{nn}=0\). We obtain
\begin{equation}
\frac{dS}{dt}
= 4\pi k_B \nu \sum_{n} n\,\dot\rho_{nn}
= 4\pi k_B \nu\,\frac{d\langle n\rangle}{dt}. \label{7.6}
\end{equation}
Hence the entropy flux carried by the escaping Proca quanta in the selected mode is
\begin{equation}
\;\frac{dS_p}{dt}\;=\;4\pi k_B\,\nu\;\dot{\bar n}_\nu\;, \label{7.7}
\end{equation}
where \(\dot{\bar n}_\nu\) is determined by \eqref{7.2} with the Proca-specific rates from Sec. \ref{sec6}. The coefficient \(4\pi\nu\) is universal: it depends only on the near-horizon mapping that fixes detailed balance, whereas the detector model, polarization content, greybody transmission, and the mass threshold \(\nu\ge m_V\) enter exclusively through \(\dot{\bar n}_\nu\). When several \(\ell\)-channels are admitted simultaneously, the total entropy flux is the sum of \eqref{7.7} over the admitted channels, each with its own \(\dot{\bar n}_\nu(\ell)\).

\subsection{Area-entropy relation with Proca modifications} \label{ssec7.2}
We now relate the entropy flux carried by the escaping Proca quanta to the change of the black hole area. Summing \eqref{7.7} over all cavity-admitted channels gives the total entropy rate
\begin{equation}
\frac{dS_p}{dt}=4\pi k_B\sum_{\nu,\ell}\nu\,\dot{\bar n}_\nu(\ell). \label{7.8}
\end{equation}
Each emitted quantum of dimensionless frequency \(\nu\) carries physical energy \(\hbar \nu_{\rm phys}\) with \(\nu_{\rm phys}=(c/r_g)\,\nu\) and \(r_g=2GM/c^2\). The radiated power at infinity is therefore
\begin{equation}
\dot E_V=\hbar\,\frac{c}{r_g}\sum_{\nu,\ell}\nu\,\dot{\bar n}_\nu(\ell)
=\hbar\,\frac{c^3}{2GM}\sum_{\nu,\ell}\nu\,\dot{\bar n}_\nu(\ell). \label{7.9}
\end{equation}

\begin{figure}
    \centering
    \includegraphics[width=\columnwidth]{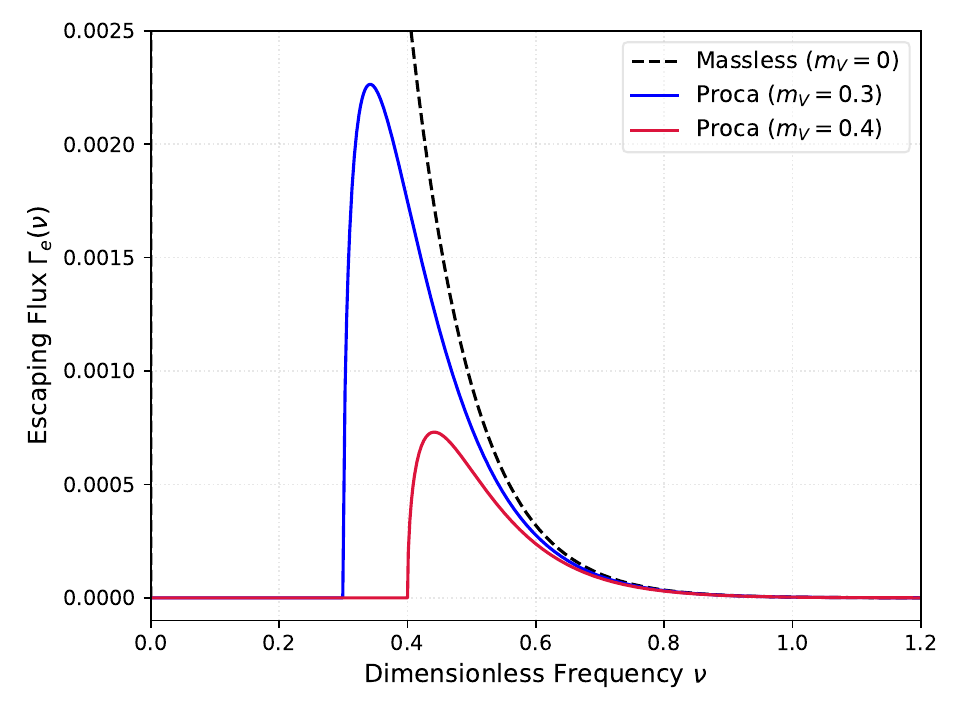}
    \caption{The escaping emission rate $\Gamma_e(\nu)$ for Proca fields with different dimensionless rest masses $m_V$. The spectra display the characteristic hard threshold at $\nu=m_V$ and the subsequent turn-on governed by phase-space availability, eventually converging to the universal massless thermal tail (dashed line) at high frequencies.}
    \label{fig3}
\end{figure}

The resulting spectral signatures of the massive vector field are summarized in Fig. \ref{fig3}. Unlike the massless case (dashed curve), which populates arbitrarily low frequencies, the Proca spectrum exhibits a hard cutoff at the rest mass $\nu=m_V$, below which the emission is strictly forbidden. Immediately above the threshold, the flux turns on smoothly, governed by the competition between the growing phase-space factor $k_\infty(\nu)$ and the exponential suppression of the thermal denominator. Crucially, in the high-frequency limit $\nu \gg m_V$, the kinematic mass effects become negligible, and all curves converge to the universal Planckian tail dictated by the HBAR entropy law. This distinct gap-plus-tail structure serves as the primary observational fingerprint of massive spin-1 acceleration radiation.

Energy conservation implies \(c^2\dot M=-\dot E_V\), and the Schwarzschild area \(A=16\pi G^2M^2/c^4\) obeys
\begin{equation}
\dot A=\frac{32\pi G^2M}{c^4}\,\dot M. \label{7.10}
\end{equation}
Combining \eqref{7.8}--\eqref{7.10} eliminates the mode sum and yields
\begin{equation}
\frac{dS_p}{dt}
= -\,\frac{k_B c^3}{4\hbar G}\,\dot A. \label{7.11}
\end{equation}
It is convenient to define the radiative area change $\dot A_V$ required to compensate for the escaping energy flux,
\begin{equation}
\dot A_V\;\equiv\;-\dot A = \frac{32\pi G^2 M}{c^4} \dot{E}_V, \label{7.12}
\end{equation}
so that the entropy-area law for the escaping massive spin-1 quanta reads
so that the entropy-area law for the escaping massive spin-1 quanta reads
\begin{equation}
\;\frac{dS_p}{dt}\;=\;\frac{k_B c^3}{4\hbar G}\,\dot A_V\;. \label{7.13}
\end{equation}
Equation \eqref{7.13} is universal: the proportionality constant depends only on \((G,\hbar,c,\,k_B)\) and the near-horizon mapping that fixed detailed balance, while all Proca-specific physics (mass threshold \(\nu\ge m_V\), polarization sums, axial/polar greybody factors, and detector realization) enters solely through \(\dot A_V\) via the net flux \(\sum_{\nu,\ell}\nu\,\dot{\bar n}_\nu(\ell)\) in \eqref{7.9}. In the massless limit \(m_V\to0\) the same relation holds, with \(\dot A_V\) computed from the corresponding threshold-free spectrum.

\subsection{Spectral signatures and limits} \label{ssec7.3}
We distill the qualitative and asymptotic features of the Proca spectrum that enter the entropy flux through \(\dot{\bar n}_\nu\) and, hence, \(\dot S_p\) in \eqref{7.7}--\eqref{7.13}. The thermal denominator is general; all structure resides in the smooth prefactor and the kinematic threshold \(\nu\ge m_V\). {\color{black}In particular, the suppression near \(\nu\simeq m_V\) is a threshold/scattering effect controlled by \(k_\infty(\nu)\) and \(\Xi_\ell(\nu)\); it is logically distinct from the large-\(\omega\) asymptotic expansion used to derive the universal near-horizon kernel.}

The Proca spectrum exhibits a sharp onset at the hard mass threshold \(\nu=m_V\). Since the asymptotic momentum is \(k_\infty(\nu)=\sqrt{\nu^2-m_V^2}\), only \(\nu\ge m_V\) contributes to the escaping flux. Near threshold the cavity-selected single-channel probability carries \(\Xi_\ell(\nu)/k_\infty(\nu)\); with the generic behavior \(\Xi_\ell(\nu)\sim k_\infty^{,2\ell+1}\) one finds \(P_{\rm exc}^{\bullet}(\nu,\ell)\propto k_\infty^{,2\ell}\). The polar sector contains \(\ell=0\) and therefore turns on most gently right above \(\nu=m_V\), while the axial sector, present only for \(\ell\ge1\), is further suppressed. At high frequencies \(\nu\gg \max{m_V,\ell}\), one has \(k_\infty\to\nu\) and \(\Xi_\ell\to1\), so the prefactor flattens to \( \propto 1/\omega^2\) and the spectral shape is controlled entirely by the universal thermal denominator \(e^{4\pi\nu}-1)^{-1}\); the spectrum displays a Wien-like peak at \(\nu=\mathcal{O}(1)\) (dimensionless) and decays exponentially thereafter.

The detector realization alters only smooth polarization-geometry weights. For the current coupling, contraction with the massive-vector projector enhances the longitudinal contribution in a way that increases smoothly as \(m_V\) decreases, but this affects only the overall spectral weight and never the thermal factor or the threshold law. For the dipole coupling, the contraction with \(F_{\mu\nu}u^\nu\) favors transverse polarizations and depends on dipole orientation; isotropic averaging yields again an order-unity, smoothly varying weight. Differences between axial and polar barriers persist close to threshold, where the polar \(\ell=0\) channel typically dominates; at larger \(\nu\) these distinctions fade as all transmissions approach unity.

The massless limit \(m_V\to0\) removes the threshold and sends \(k_\infty\to\nu\), returning the scalarlike normalization \(1/\nu\) (up to polarization weights). The dipole model passes smoothly to the standard electromagnetic response. {\color{black}These observations also clarify the semirealistic physical meaning of the Proca extension. The new ingredients are not limited to extra algebraic structure: the massive vector field predicts a hard low-frequency gap, a polarization/channel hierarchy close to threshold, and coupling-dependent longitudinal/transverse weights, all of which are absent in the scalar HBAR problem. In a cavity-resolved or detector-based realization, these quantities would be the direct observables distinguishing massive spin-1 acceleration radiation from its scalar or Maxwell counterparts. The present analysis therefore identifies which parts of the signal are universal consequences of near-horizon kinematics and which parts encode the genuinely new physics of a Proca field.} The current model, while formulated naturally for Proca, still yields a smooth limit at the level of observable probabilities and rates. Finally, all results here rest on the large-\(\omega\) expansion that produced the universal near-horizon integral; corrections scale as \(\mathcal O(\nu/\omega)\), reshaping only the smooth prefactor and leaving intact the Planckian denominator and detailed balance. Consequently, the entropy flux \(\dot S_p=4\pi k_B\sum_{\nu,\ell}\nu\dot{\bar n}_\nu(\ell)\) inherits the same signatures: delayed onset at \(\nu\approx m_V\), gentle turn-on dominated by polar \(\ell=0\), and a universal thermal tail at higher frequencies.

These signatures (sharp threshold, channel-dependent turn-on, polarization-weighted prefactors, and a universal thermal tail) are the distinctive Proca imprints in the HBAR entropy law \eqref{7.13}.

\section{Consistency checks and theoretical subtleties} \label{sec8}
We assess the reliability and scope of the large-\(\omega\) expansion used to extract the universal thermal factor, clarify how the Proca constraint replaces gauge freedom and how it propagates through the mode expansion and detector couplings, and frame our near-horizon derivation in the language of Unruh/DeWitt response theory. The takeaways are: (i) the Planckian denominator and detailed balance are controlled by the local Rindler mapping and are insensitive to smooth prefactors; (ii) the massive vector has three physical polarizations with \(\nabla_\mu A^\mu=0\) enforced dynamically, so no gauge subtleties pollute the result; (iii) the calculation is the curved-space analogue of a detector crossing a Rindler horizon, with the same KMS periodicity responsible for the thermal factor.

\subsection{Large-\texorpdfstring{\(\omega\)}{} expansion and validity} \label{ssec8.1}
The first-order amplitudes reduce to a near-horizon Laplace-type integral,
\begin{equation}
\mathcal A_\bullet(\nu,\ell,m,\lambda)\;=\;\int d\tau\;\mathcal G_\bullet(\tau,\nu,\ell,m,\lambda)\,
e^{\,i\Phi(\tau,\nu)}. \label{8.1}
\end{equation}
Near the horizon, the phase is
\begin{equation}
\Phi(\tau,\nu)=\omega \tau+\nu\,\left[t(\tau)-r_*(\tau)\right]+\Phi_0(\nu), \label{8.2}
\end{equation}
with \(\Phi_0(\nu)\) \(\tau\)-independent (absorbing matching constants such as \(\nu-k_\infty\)). Setting \(y\equiv \tau_h-\tau>0\) and using \(t-r_*=-2\ln y + C_h + \mathcal O(y)\), we obtain the universal integral
\begin{equation}
\mathcal I(\nu,\omega)\;\equiv\;\int_0^\infty dy\,y^{-2i\nu}\,e^{-i\omega y}
\;=\;(i\omega)^{-(1-2i\nu)}\Gamma(1-2i\nu), \label{8.3}
\end{equation}
so that
\begin{equation}
|\mathcal I(\nu,\omega)|^2=\frac{4\pi\,\nu}{\omega^2}\,\frac{1}{e^{4\pi\nu}-1}. \label{8.4}
\end{equation}

{\color{black}The large-\(\omega\) control arises because the integrand oscillates rapidly on a scale \(y\sim \omega^{-1}\). Provided the envelope is smooth on this scale, we may expand both the smooth prefactor and the regular part of the phase about the horizon. Writing
\begin{equation}
\mathcal G_\bullet(\tau_h-y,\nu,\cdot)\,e^{\,i\nu \mathcal O(y)}
=\sum_{n=0}^{\infty} c_n^{(\bullet)}(\nu,\cdot)\,y^n, \label{8.5}
\end{equation}
the amplitude becomes
\begin{align}
\mathcal A_\bullet
&=\sum_{n=0}^{\infty} c_n^{(\bullet)}(\nu,\cdot)\,\mathcal I_n(\nu,\omega), 
\nn
\mathcal I_n(\nu,\omega)&\equiv \int_0^\infty dy\,y^{n-2i\nu}e^{-i\omega y}. \label{8.6}
\end{align}
These moments are evaluated exactly as
\begin{equation}
\mathcal I_n(\nu,\omega)
=\frac{\Gamma(n+1-2i\nu)}{(i\omega)^{\,n+1-2i\nu}}
=\mathcal I(\nu,\omega)\,\frac{(1-2i\nu)_n}{(i\omega)^n}, \label{8.7}
\end{equation}
where \((a)_n\) is the Pochhammer symbol. Therefore
\begin{equation}
\mathcal A_\bullet
=\mathcal I(\nu,\omega)\left[
c_0^{(\bullet)}
+\frac{(1-2i\nu)c_1^{(\bullet)}}{i\omega}
+\mathcal O\!\left(\frac{1+\nu^2}{\omega^2}\right)\right], \label{8.8}
\end{equation}
and, correspondingly,
\begin{equation}
P_\bullet(\nu,\ell)
=P_\bullet^{(0)}(\nu,\ell)\left[
1+\mathcal O\!\left(\frac{1+|\nu|}{\omega}\right)\right]. \label{8.9}
\end{equation}
Thus the asymptotic parameter is \(1/\omega\), while the emitted-mode frequency \(\nu\) enters only through smooth polynomial coefficients multiplying the same universal kernel \(\mathcal I(\nu,\omega)\).

This makes explicit that the low-frequency suppression discussed in Sec. \ref{ssec7.3} is not produced by a breakdown of the large-\(\omega\) expansion. Rather, it comes from the escaping-mode prefactor
\begin{equation}
P_{\mathrm{exc,esc}}^\bullet(\nu,\ell)\propto
\frac{\Theta(\nu-m_V)}{k_\infty(\nu)}\,\Xi_\ell(\nu)\,
\frac{1}{e^{4\pi\nu}-1}, \label{8.10}
\end{equation}
with \(k_\infty(\nu)=\sqrt{\nu^2-m_V^2}\) and \(\Xi_\ell(\nu)\to0\) as \(\nu\to m_V^+\). In particular, if \(\Xi_\ell(\nu)\sim k_\infty^{2\ell+1}\) near threshold, then \(P_{\mathrm{exc,esc}}^\bullet(\nu,\ell)\propto k_\infty^{2\ell}\), so the turn-on at low \(\nu\) is a threshold/greybody effect rather than a failure of the \(1/\omega\) expansion.

The universal Planckian denominator and the detailed-balance ratio originate from the common factor \(\mathcal I(\nu,\omega)\), i.e. from the logarithmic near-horizon mapping of \(t-r_*\). Subleading terms only renormalize the smooth channel-dependent prefactor and therefore affect absolute rates and relaxation times, not the leading geometric thermal kernel. The regime of validity is
\begin{equation}
\omega\,\Delta\tau_{\rm cav}\gg 1,
\qquad
\Delta\nu\ll \min\{\nu,\omega\},
\qquad
\frac{1+|\nu|}{\omega}\ll 1. \label{8.11}
\end{equation}}

Two further checks secure the approximation. First, the contribution from \(y\gtrsim\omega^{-1}\) is exponentially suppressed by destructive interference. Since the cavity bottom is taken to \(r_b\to 1^+\), the dominant \(y\)-region is entirely inside the cavity for large enough \(\omega\). Quantitatively, if \(\Delta\tau_{\rm cav}\) is the proper-time span inside the cavity, we require
\begin{equation}
   \omega\,\delta\tau_{\rm cav}\gg 1 \quad\Rightarrow\quad
   \text{single-pass, near-horizon dominance.} \label{8.8}
\end{equation}
Second, the asymptotic momentum \(k_\infty(\nu)=\sqrt{\nu^2-m_V^2}\) enters only through the smooth matching constants and the unit-flux normalization \(1/\sqrt{k_\infty}\) at infinity. The greybody factor \(\Xi_\ell(\nu)\to 0\) as \(k_\infty\to0^+\) ensures that the composite prefactor \(\Xi_\ell/k_\infty\) is finite or vanishing near \(\nu\to m_V^+\); the Planck denominator is unaffected.

Finally, the finite top mirror at \(r_t\) selects a narrow frequency window. Provided the window width \(\Delta\nu\) obeys \(\Delta\nu\ll \nu\) (so that \(\mathcal G_{\bullet h}\) and \(\Xi_\ell\) are effectively constant across it) and \(\Delta\nu\ll \omega\) (so that the phase analysis is unchanged), the replacement \(\sum_n\to\int d\nu\,\rho(\nu)\) with \(\rho(\nu)\propto \nu/k_\infty\) remains accurate.

\subsection{Gauge and constraints} \label{ssec8.2}
The Proca theory is massive and therefore has no gauge symmetry. The covariant Lorenz condition \(\nabla_\mu A^\mu=0\) follows dynamically from the equations of motion, leaving three propagating degrees of freedom. This structure percolates cleanly through mode expansion and detector couplings. Asymptotically, the three polarizations satisfy \(k\!\cdot\!\varepsilon^{(\lambda)}=0\), and the projector is \cite{tong2014quantum}
\begin{equation}
    \sum_{\lambda=1}^{3}\varepsilon^{(\lambda)}_\mu \varepsilon^{(\lambda)}_\nu = -\eta_{\mu\nu}+\frac{k_\mu k_\nu}{m_V^2}. \label{8.9}
\end{equation}
Near the horizon, the axial/polar split maps to three independent master variables with plane wave behavior \(e^{-i\nu(t-r_*)}\); the constraint does not modify the universal phase.

For the current (monopole) coupling, the interaction \(g_V m(\tau)u^\mu A_\mu\) respects the Proca constraint because \(u^\mu\) is nondynamical and confined to the worldline. The polarization sum entering \(\sum_\lambda|u\!\cdot\!\varepsilon^{(\lambda)}|^2\) is
\begin{equation}
    -\,u\!\cdot\!u+\frac{(u\!\cdot\!k)^2}{m_V^2}
    =\,-1+\frac{\Omega^2}{m_V^2}, \label{8.10}
\end{equation}
with \(\Omega=-k\!\cdot\!u\). This factor is positive and large near the horizon where \(\Omega\gg m_V\), and it multiplies (but cannot alter) the thermal denominator.

For the physical dipole coupling, with \(L_{\rm int}=-d^\mu F_{\mu\nu}u^\nu\) and \(u\!\cdot\!d=0\), the contraction
\begin{equation}
    E_\mu\equiv F_{\mu\nu}u^\nu = i\left[k_\mu(A\!\cdot\!u)-(k\!\cdot\!u)\,A_\mu\right] \label{8.11}
\end{equation}
automatically satisfies \(u^\mu E_\mu=0\). Both pieces contribute at the amplitude level; after summing polarizations the weight is built from contractions with the projector in \eqref{8.9}. Orientation averaging of \(d^\mu\) in the atom's rest frame produces a smooth, \(\mathcal O(1)\) function of \(\nu\).

Finally, for the massless limit, as \(m_V\to0\), the projector \eqref{8.9} becomes singular, but all observables remain finite: the combination of unit-flux normalization, \(\Xi_\ell(\nu)\), and the longitudinal/transverse balance yields a smooth limit in probabilities and rates. Physically, the dipole model tends to the familiar electromagnetic response; the current model, while not gauge-invariant in the \(m_V=0\) theory, approaches the same observables because the longitudinal contribution decouples as \(\nu\) grows and \(\Xi_\ell\to1\).

In all cases mentioned above, the constraint only shapes prefactors. The near-horizon analytic structure that enforces the \(t-r_*\) phase, and hence the thermal factor, remains untouched.

\subsection{Comparison with Unruh/DeWitt logic} \label{ssec8.3}
Our calculation is the black hole (Killing/Rindler) analogue of the Unruh-DeWitt (UDW) detector response. Three points make the connection precise. First, in the UDW setting, a uniformly accelerated detector with proper acceleration \(a\) sees a thermal spectrum with temperature \(T=a/(2\pi)\), traced to the KMS periodicity of the Wightman function in imaginary proper time. Here, along the free fall trajectory, the near-horizon mapping
\begin{equation}
   t(\tau)-r_*(\tau)=-\,2\ln(\tau_h-\tau)+C_h+\cdots \label{8.12}
\end{equation}
acts as a local Rindler time. The logarithm produces a branch cut whose discontinuity across imaginary \(2\pi\) shifts enforces \cite{Unruh:1976db,Hartle:1976tp}
\begin{equation}
   \frac{P_{\rm exc}(\nu)}{P_{\rm abs}(\nu)}=e^{-4\pi\nu}, \label{8.13}
\end{equation}
exactly the KMS ratio with surface gravity \(\kappa=\frac{1}{2}\) in our dimensionless units. This universality explains why spin, mass, and detector realization cannot change the thermal denominator. Second, in the Boulware vacuum, the rotating term \(\propto a_{\nu\ell m\lambda}\sigma_{eg}\) vanishes on \(|0\rangle\), while the counterrotating term \(\propto a^\dagger_{\nu\ell m\lambda}\sigma_{eg}\) survives. This is the curved-space counterpart of the UDW vacuum excitation via antiresonant terms, and is the microscopic origin of the emission channel that populates the mode in our master equation. Third and finally, a static detector at fixed \(r\) in the Unruh picture samples a local temperature \(\propto 1/\sqrt{f(r)}\) \cite{Tjoa:2020eqh}. Our atoms are not static; they free fall across the near-horizon region. The observable is the escaping flux at \(\mathscr I^+\), obtained by sewing the near-horizon thermal kernel (which fixes detailed balance) to the global scattering problem [which provides \(\Xi_\ell(\nu)\) and the threshold]. The result is the same thermal ratio \eqref{8.13} multiplied by smooth, channel-dependent prefactors. That is why the entropy law
\begin{equation}
   \frac{dS_p}{dt}=4\pi k_B\sum_{\nu,\ell}\nu\,\dot{\bar n}_\nu(\ell) \label{8.14}
\end{equation}
and the area-entropy relation
\begin{equation}
   \frac{dS_p}{dt}=\frac{k_B c^3}{4\hbar G}\,\dot A_V \label{8.15}
   \end{equation}
hold independent of vector mass and polarization content: the Proca details flow only into \(\dot{\bar n}_\nu\) through thresholds and greybody transmissions \cite{Bekenstein:1973ur,Hawking:1975vcx,Bardeen:1973gs}.

The near-horizon analytic structure supplies the same KMS kernel that underlies the Unruh/DeWitt response; our curved-space addition is the propagation (and selective leakage) of those quanta to infinity through polarization-dependent potentials. This factorization cleanly separates the universal thermal factor from the model-dependent prefactor, validating the use of the master equation and the HBAR-style entropy law for the massive spin-1 case.

\section{Conclusion} \label{sec9}
In this work, we extended Scully \textit{et al.}'s acceleration-radiation framework \cite{Scully:2017utk} from a scalar field to a massive spin-1 (Proca) field for atoms freely falling into a Schwarzschild background, treating both detector realizations: a charged-monopole current coupling and a physical electric-dipole coupling. For both detector models, the near-horizon, counterrotating amplitude reduces to the same Laplace integral, giving the universal spectral kernel in \eqref{5.12} and the detailed-balance ratio in \eqref{5.18}. This arises from the local Rindler mapping of \(t-r_*\) and is insensitive to spin, mass, polarization, or detector realization.

Absolute excitation probabilities acquire smooth prefactors and a hard mass threshold via \(k_\infty(\nu)\), greybody transmission \(\Xi_\ell(\nu)\), and polarization/geometry weights. For instance, see the general structure \eqref{5.14} and its current/dipole specializations \eqref{5.15}--\eqref{5.16}. Near threshold, polar \(\ell=0\) turns on most gently, while axial channels are more strongly suppressed.

Converting probabilities into escaping emission/absorption rates yields the detailed-balance relation \eqref{6.8}. The single-mode master equation relaxes to the geometric steady state with mean occupancy given in \eqref{6.13}. Detector, mass, and polarization only set relaxation speeds and flux magnitudes, not the equilibrium ratio.

The master equation implies the universal mode-by-mode entropy flux rule \eqref{7.7}. Combined with energy balance and the Schwarzschild area law, this gives the HBAR-style area-entropy relation \eqref{7.13}. The proportionality constant is universal; all Proca specifics enter only through the radiative area change via the net flux.

{\color{black}The Proca spectrum displays a sharp onset at the mass threshold, channel-dependent near-threshold scaling governed by \(\Xi_\ell(\nu)\) and \(k_\infty(\nu)\), a high-frequency regime where \(\Xi_\ell\to1\) and the Planckian tail dominates, and polarization-sensitive weights distinguishing current from dipole couplings. Taken together, these imply a characteristic gap-plus-tail spectral profile with longitudinal/transverse and axial/polar discrimination that is absent in the scalar problem. This is the concrete physical payoff of the Proca extension: it identifies the observable spin-1 signatures that survive on top of the same universal near-horizon thermal kernel. We emphasize, however, that the present paper provides the analytic detector-theory template for those signatures rather than a complete astrophysical detectability analysis.}

{\color{black}The large-\(\omega\) expansion confines corrections to the smooth prefactor, with relative size controlled by \(\mathcal O((1+|\nu|)/\omega)\) in the cavity-selected regime, leaving the universal thermal kernel (V.13) and the leading detailed-balance structure intact.}. The Proca constraint \(\nabla_\mu A^\mu=0\) replaces gauge freedom cleanly, and the massless limit is smooth at the level of probabilities/rates. The factorization into a universal kernel times a smooth prefactor justifies the master-equation treatment and the entropy law for massive spin-1 quanta.

{\color{black}For future directions, explicit axial/polar greybody factors, rotating backgrounds (superradiance), alternative exterior states (Unruh/Hartle-Hawking), and detector-engineering for longitudinal/transverse isolation would sharpen the semirealistic phenomenology developed here. In particular, they would turn the present analytic signatures:threshold gap, channel-dependent turn-on, and polarization-sensitive response, which are into a more quantitative bridge between the HBAR detector framework and massive-vector scenarios of current interest.}

\acknowledgments
R. P. and A. \"O. would like to acknowledge networking support of the COST Action CA21106 - COSMIC WISPers in the Dark Universe: Theory, astrophysics and experiments (CosmicWISPers), the COST Action CA22113 - Fundamental challenges in theoretical physics (THEORY-CHALLENGES), the COST Action CA21136 - Addressing observational tensions in cosmology with systematics and fundamental physics (CosmoVerse), the COST Action CA23130 - Bridging high and low energies in search of quantum gravity (BridgeQG), and the COST Action CA23115 - Relativistic Quantum Information (RQI) funded by COST (European Cooperation in Science and Technology). R. P. and A. \"O. would also like to acknowledge the funding support of SCOAP3 (Switzerland). A. \"O. also thanks EMU, TUBITAK, ULAKBIM (Turkiye).

\bibliography{refs}

\end{document}